\documentclass[10pt,journal,compsoc]{IEEEtran}

\ifCLASSOPTIONcompsoc
  \usepackage[nocompress]{cite}
\else
  \usepackage{cite}
\fi

\usepackage{cite}
\usepackage{amsmath,amsfonts}
\usepackage{algorithmic}
\usepackage{xspace}
\usepackage{graphicx}
\usepackage{textcomp}
\usepackage{xcolor}
\usepackage{tabularx}
\usepackage{array}  
\usepackage{booktabs}
\usepackage{hyperref}
\usepackage{caption}
\usepackage{amsmath}
\usepackage{enumitem}
\usepackage{subcaption}
\usepackage{listings}
\usepackage{siunitx}
\usepackage{multirow}
\usepackage{makecell}
\usepackage{url}
\usepackage{enumitem}
\usepackage{refcount}

\usepackage[absolute,overlay]{textpos}
\usepackage{tikz}
\usepackage{xcolor}
\usepackage{calc}

\newcounter{findingctr}

\newcommand{\finding}[2]{\refstepcounter{findingctr}\emph{#1:\label{box:#2}}}

\newlength{\boxw}
\newlength{\boxh}
\newlength{\shadowsize}
\newlength{\boxroundness}
\newlength{\tmpa}
\newsavebox{\shadowblockbox}

\newenvironment{findingenv}[2]%
{\vspace{0.2cm}\noindent
\begin{lrbox}{
\shadowblockbox
}
\begin{minipage}{.98\columnwidth}
\finding{#1}{#2}~}%
{\end{minipage}\end{lrbox}%
\settowidth{\boxw}{\usebox{\shadowblockbox}}%
\settodepth{\tmpa}{\usebox{\shadowblockbox}}%
\settoheight{\boxh}{\usebox{\shadowblockbox}}%
\addtolength{\boxh}{\tmpa}%
\begin{tikzpicture}
\addtolength{\boxw}{\boxroundness * 2}
\addtolength{\boxh}{\boxroundness * 2}

\foreach \x in {0,.05,...,1}
{
\setlength{\tmpa}{\shadowsize * \real{\x}}
\fill[xshift=\shadowsize - 1pt,yshift=-\shadowsize + 
1pt,black,opacity=.04,rounded corners=\boxroundness] 
(\tmpa, \tmpa) rectangle +(\boxw - \tmpa - \tmpa, \boxh - \tmpa - 
\tmpa);
}

\filldraw[fill=white!50, draw=black!80, rounded corners=\boxroundness] (0, 
0) rectangle (\boxw, \boxh);
\draw node[xshift=\boxroundness,yshift=\boxroundness,inner sep=0pt,outer 
sep=0pt,anchor=south west] (0,0) {\usebox{\shadowblockbox}};
\end{tikzpicture}\vspace{0cm}%
}

\newcommand{\revision}[1]{{~#1}\xspace}

\newcommand{\intellimerge}{\texttt{IntelliMerge}\xspace}
\newcommand{\ourtool}{\texttt{RefMerge}\xspace}
\newcommand{\git}{Git\xspace}
\newcommand{\sample}{2,001\xspace}

\newcommand{\checkNum}[1]{{#1}\xspace}

\def\BibTeX{{\rm B\kern-.05em{\sc i\kern-.025em b}\kern-.08em
    T\kern-.1667em\lower.7ex\hbox{E}\kern-.125emX}}
\begin{document}

\title{\revision{
Operation-based Refactoring-aware Merging: An Empirical Evaluation}}

\author{
    Max Ellis, 
    Sarah Nadi, 
    Danny Dig
}


\IEEEtitleabstractindextext{
\begin{abstract}
Dealing with merge conflicts in version control systems is a challenging task for software developers.
Resolving merge conflicts is a time-consuming and error-prone process, which distracts developers from important tasks. 
Recent work shows that refactorings are often involved in merge conflicts and that refactoring-related conflicts tend to be larger, making them harder to resolve.
In the literature, there are two refactoring-aware merging techniques that claim to automatically resolve refactoring-related conflicts; however, these two techniques have never been empirically compared. 
\revision{
In this paper, we present \ourtool, a rejuvenated Java-based design and implementation of the first technique, which is an operation-based refactoring-aware merging algorithm.
We compare \ourtool to \git and the state-of-the-art graph-based refactoring-aware merging tool, \intellimerge, on 2,001 merge scenarios with refactoring-related conflicts from 20 open-source projects. 
We find that \ourtool resolves or reduces conflicts in 497 (25\%) merge scenarios while increasing conflicting LOC in only 214 (11\%) scenarios.
On the other hand, we find that \intellimerge resolves or reduces conflicts in 478 (24\%) merge scenarios but increases conflicting LOC in 597 (30\%) merge scenarios. We additionally conduct a qualitative analysis of the differences between the three merging algorithms and provide insights of the strengths and weaknesses of each tool.
We find that while \intellimerge does well with ordering and formatting conflicts, it struggles with class-level refactorings and scenarios with several refactorings.
On the other hand, \ourtool is resilient to the number of refactorings in a merge scenario, but we find that \ourtool introduces conflicts when inverting move-related refactorings.
}


\end{abstract}

}

\maketitle

\IEEEraisesectionheading{\section{Introduction}\label{sec:introduction}}

\revision{
Version control systems (VCS) play a crucial role in enabling developers to collaborate on software projects. 
Whether developers are working on the same branch~\cite{bird}, using branch-based development~\cite{Phillips}, or using pull requests to contribute changes from their external forks~\cite{externalForks}, integration issues can arise when they push their changes to the repository. When two\footnote{Note that in this paper, we focus on the common practice of merging of changes from two versions of the code, and do not consider what is often referred to as \textit{octopus merges} when more than two branches/versions are involved.} developers try to contribute different changes to the same part of the code, a VCS reports a \textit{merge conflict}.
Based on analyzing four open-source projects, previous work found that merge conflicts occurred up to 19\% of the time and could sometimes take several days to resolve~\cite{cassandra}.
As our own analysis of 20 open-source projects shows, this percentage can vary significantly per project (3\% - 55\%), with a median of 16\% (See Table~\ref{tab:conflict-scenarios}).
Even worse, existing merge tools cannot detect every merge conflict; such conflicts might not be discovered until building or testing and may even be released in software products, causing unexpected behavior~\cite{Ahmed, cassandra}.
Thus, overall, while merge conflicts are moderately frequent, they are a burden when they occur.
A recent practitioner survey shows that developers spend time trying to understand and resolve conflicts and that current support tools do not meet all their conflict-resolution needs~\cite{mckee2017software}.

A common issue with merge conflicts is that most modern version control systems, such as Git~\cite{git}, Mercurial~\cite{mercurial}, or SVN~\cite{svn}, treat all stored artifacts as plain text and merge files line by line.
When two different changes happen to the same line of code, a textual line-based merging tool (often referred to as an \textit{unstructured merge tool}~\cite{Berlin}) will report a conflict since it cannot automatically decide which change to choose.
However, a tool that understands the nature of the code change that occurred may be able to resolve the conflict~\cite{MahmoudAndroid2018,Apel}.
For example, \textit{refactorings} are code changes that 
modify the structure of the code to improve its readability or maintainability without altering its observable behavior~\cite{Fowler}. 
Refactorings are one example of a code change with well-defined semantics that an automated merge-conflict resolution tool can understand and automatically resolve~\cite{SANER,MahmoudAndroid2018, IntelliMerge, MolhadoRef}.
For example, if Bob refactors method \texttt{foo} by moving it from one class to another on one branch while Alice, on another branch, adds a line of code to \texttt{foo}'s body, an unstructured merging tool will report a merge conflict because Bob and Alice changed the same lines of code. However, a merge tool that is aware of the semantics behind these refactoring changes can resolve this conflict by adding the new line of code to \texttt{foo}'s new location.
Thus, understanding the semantics of refactorings could avoid unnecessary merge conflicts and save developers' time.
A recent study found that 15 of more than 70 known refactorings are involved in 22\% of merge conflicts and tend to result in larger conflicts~\cite{SANER}. The considerable portion of merge conflicts that refactorings complicate motivates the need for automated merging tools that can handle refactorings.
}

\revision{While there are several research efforts that work on understanding the structure of underlying code to automate more merge-conflict resolutions~\cite{Apel, struct1, Cavalcanti, JDime, lessenich2015balancing, AutoMerge, Seibt}, there are mainly two efforts that specifically focus on refactorings. 
The first is by Dig et al.~\cite{MolhadoRef} that proposes an \textit{operation-based refactoring-aware merging technique}, MolhadoRef. 
At a high level, given two branches to be merged, MolhadoRef first inverts refactorings on both branches, textually merges the refactoring-free version of the code, and then replays the refactorings on the merged code. 
Their evaluation, based on one project, shows a 97\% reduction of merge conflicts.

The second approach by Shen et al.~\cite{IntelliMerge} is a graph-based refactoring-aware merging approach implemented in \intellimerge. \intellimerge converts code on both branches to graphs, and performs a graph-based three-way merge (i.e., considering the common ancestor of both branches too) where it tries to match nodes across the three versions. 
This node matching is based on a set of predefined rules that are meant to capture refactoring semantics along with a similarity score threshold. 
The authors evaluate \intellimerge on 10 projects and report 88\% and 90\% precision and recall, respectively, when compared to the resolution committed by developers.

While both approaches show promise in their evaluation, they each have limitations. To begin, the premise of MolhadoRef is that if the version-control history records \textit{operations} (i.e., the types of code changes that occur instead of simple textual changes), then we can leverage these refactoring operations in the history to resolve conflicts. 
To that end, MolhadoRef relies on developers using the researchers' operation-based version control system, Molhado, preventing the approach from being used on modern version control systems. 
Furthermore, MolhadoRef's implementation supported only six refactorings, none of which are complex refactorings such as \textit{Extract Method} and \textit{Inline Method}, which the authors of \intellimerge~\cite{IntelliMerge} argued limit operation-based techniques due to the difficulty of inverting them. 
Finally, while their evaluation shows promise, it was limited to their own repository, thus MolhadoRef's feasibility in practice is unknown, especially since there is no publicly available implementation of their approach. Even if there was, MolhadoRef's reliance on Molhado makes replicating it on real-world projects impossible.

On the other hand, \intellimerge relies on a similarity score for detecting refactorings, which has been argued in the literature to misidentify or miss refactorings~\cite{RefactoringMiner1}. 
This can lead to introducing additional conflicts, unexpected merges, or missing conflicts.
In addition, \intellimerge does not consider how refactorings on each branch will interact with each other.
Furthermore, it has been evaluated only w.r.t. resolving more conflicts using imprecise metrics and the evaluation does not analyze whether the tool's resolutions are correct or whether there were missed conflicts that \intellimerge did not detect. 
Finally, while criticizing operation-based merging, the authors never performed an evaluation comparing the two approaches.
Given the merit of solving merge conflicts when refactorings are involved, we believe that a direct comparison will shed light on the strengths and weaknesses of these techniques.
Such insights can help push the state of the art of refactoring-aware merging techniques further.
}

\begin{figure*}[t]
	\centering
	\begin{subfigure}{.45\textwidth}
		\centering
		\includegraphics[width=\textwidth]{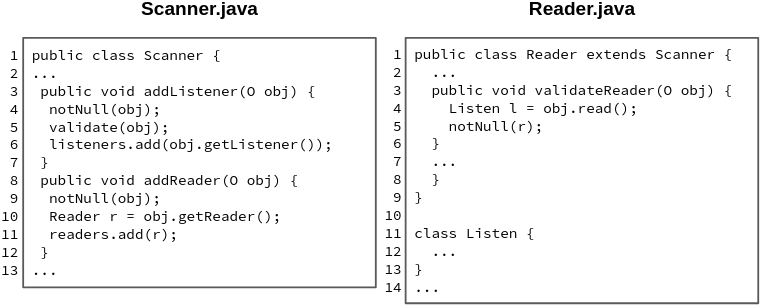}
		\caption{Base commit}
		\label{base}
	\end{subfigure}\\
		\begin{subfigure}{.45\textwidth}
		\centering
		\includegraphics[width=\textwidth]{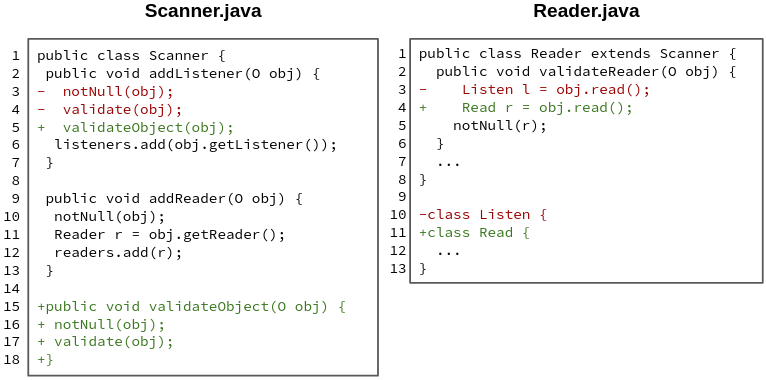}
		\caption{Left parent}
		\label{left}
	\end{subfigure}
		\begin{subfigure}{.45\textwidth}
		\centering
		\includegraphics[width=\textwidth]{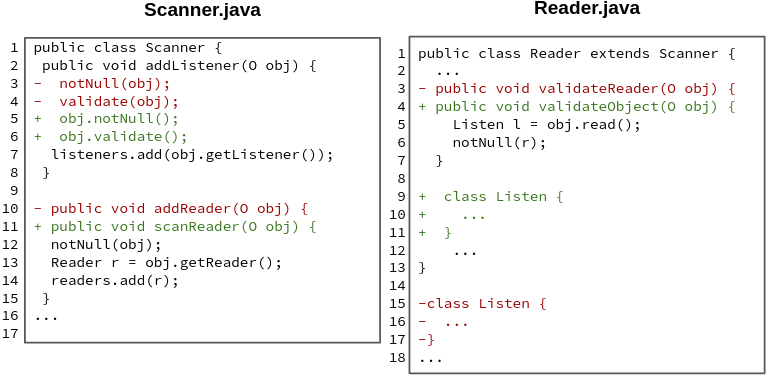}
		\caption{Right parent}
		\label{right}
	\end{subfigure}
	\begin{subfigure}{.45\textwidth}
		\centering
		\includegraphics[width=\textwidth]{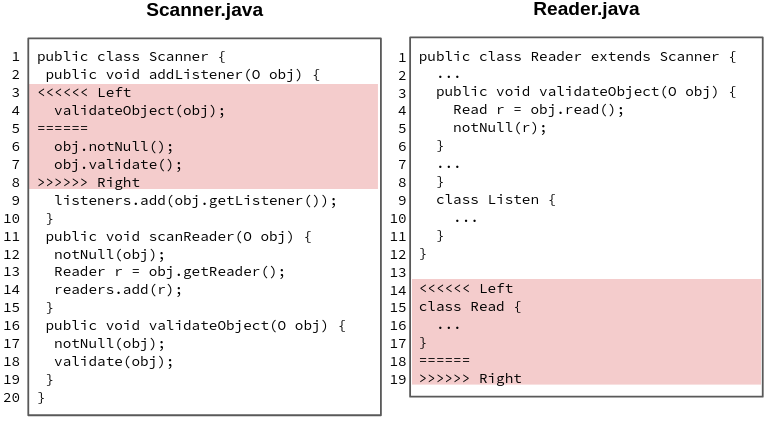}
		\caption{Merge result for Git}
		\label{git}
	\end{subfigure}
		\begin{subfigure}{.45\textwidth}
		\centering
		\includegraphics[width=\textwidth]{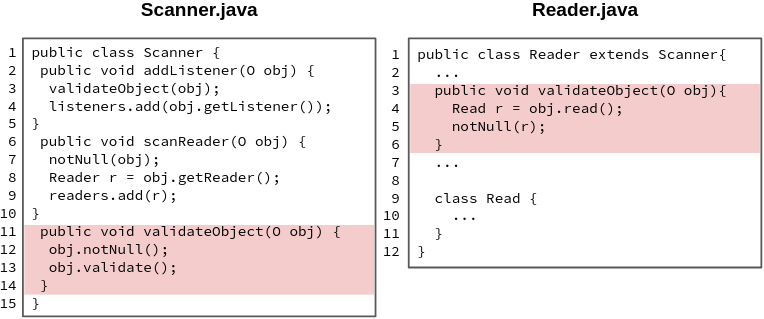}
		\caption{Ideal merge result}
		\label{refMerge}
	\end{subfigure}
	
\caption{The three versions (base, left, and right) of code from Scanner.java and Reader.java, as well as the results merged by \git and an ideal merge tool.}
\label{example}
\vspace{-4mm}
\end{figure*}

\revision{
{To enable the comparison of these two techniques, this paper has two goals. Our first goal is to rejuvenate Dig et al.'s operation-based refactoring-aware merging technique~\cite{MolhadoRef}.}
This requires a re-design of the technique to enable it to work with modern VCSs, such as git. We implement our rejuvenated operation-based refactoring-aware merging technique in \ourtool. 
While \ourtool follows the same approach of reverting and replaying refactorings, there are several novelties that differentiate \ourtool from MolhadoRef: 
(1) Whereas MolhadoRef relies on a research-based version control system, \ourtool is designed to work directly on top of \git, since it is the most popular version control system used by practitioners~\cite{brindescu2014centralized}, 
(2) \ourtool supports 17 refactoring types (instead of MolhadoRef's six), including \textit{Extract Method} and \textit{Inline Method} which were argued to not have an inverse refactoring~\cite{IntelliMerge},
(3) To detect refactorings in \git history, \ourtool uses the state-of-the-art refactoring detection tool, RefactoringMiner~\cite{RefMiner}, 
(4) \ourtool avoids checking for circular dependencies by simplifying and combining refactorings upon detection, and finally 
(5) We evaluate \ourtool on a large scale to determine the feasibility of operation-based refactoring-aware merging in practice. 

Our second goal is to compare the two refactoring-aware merging techniques on real-world projects that use \git as their version control system, since it is the most popular version control system used by practitioners~\cite{brindescu2014centralized}. To that end, we perform the first large-scale comparison of operation-based merging and graph-based merging techniques.
In summary, this paper makes the following contributions:

\begin{itemize}[leftmargin=*]

    \item An open-source design and implementation~\cite{artifact} of operation-based refactoring-aware merging, \ourtool, built on top of \git and which covers 17 refactoring types, including two complex refactorings that complicate conflicts~\cite{SANER} and were proposed to be difficult for operation-based merging~\cite{IntelliMerge}, \textit{Extract Method} and \textit{Inline Method}.
    
    \item A large-scale \textit{quantitative} comparison of the effectiveness of operation-based refactoring implemented in \ourtool versus graph-based refactoring implemented in \intellimerge. Our evaluation includes \sample merge scenarios from 20 open-source projects.
    
    \item A systematic \textit{qualitative} comparison of the strengths and weaknesses of both techniques through a manual analysis of their results across a sample of \checkNum{50} merge scenarios.
    
    \item A discussion of how refactoring-aware merging can be improved based on the identified strengths and weaknesses of the two techniques.
    
\end{itemize}

}

\revision{
Our evaluation results show that while \intellimerge reduces the number of refactoring conflicts a developer needs to deal with, graph node matching errors and the reliance on a similarity score cause \intellimerge to highly increase the number of false positives and false negatives. On the other hand, \ourtool is able to reduce the number of false positives while eliminating false negatives. However, \ourtool sometimes introduces conflicts while inverting move-related refactorings. Our findings shed light on how both refactoring-aware approaches can be improved, and we recommend adding support for more refactorings with operation-based merging.
Our complete replication package is available online~\cite{artifact}.
}


\section{Background and Motivating Example}

To introduce the terms we use, we briefly describe how merging works in \git. We also provide an example to motivate the need for refactoring-aware merging techniques.

\subsection{Software Merging in \git}

\begin{figure*}[t!]
\includegraphics[width=\textwidth]{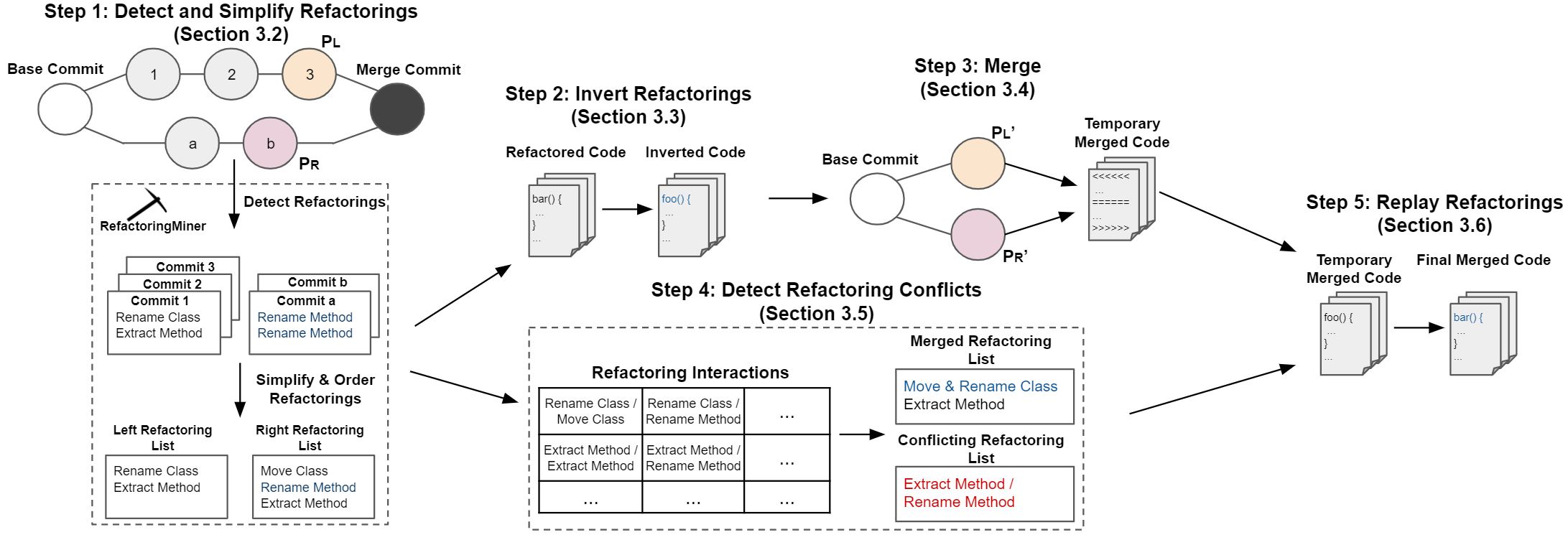}
\caption{An overview of RefMerge's merging algorithm}
\label{overviewDiagram}
\vspace{-4mm}
\end{figure*}

A \textit{merge scenario} occurs when developers using \git need to integrate changes they separately worked on in different branches.
The merge tools that are commonly utilized by VCSs such as \git use three-way merging techniques~\cite{Mens}. 
In \textit{three-way merging}, two versions of the software are merged by making use of these versions' \textit{common ancestor}, which is the common version of the code the two versions originated from before they started diverging.
When merging two branches, \git attempts to merge the most recent commit on each branch, which we refer to as the \textit{parent commits}, using the common ancestor of these commits, which we refer to as the \textit{base commit}.
The result of the merge is stored in a \textit{merge commit}.
An example of a commit history leading to a merge commit is shown at the top left corner of Figure~\ref{overviewDiagram}.

A \textit{conflicting merge scenario} is one where a merge tool is not able to automatically merge the changes from the two versions being integrated.
\git reports the conflicting locations by annotating them with \texttt{<<<}, \texttt{===}, and \texttt{>>>} markers. We call these regions \textit{conflict blocks}.
When a file contains at least one conflict block, we refer to the file as a \textit{conflicting file}. We refer to the lines within the conflict block as conflicting lines of code, or \textit{conflicting LOC}.
For example, Figure~\ref{git} shows two conflicting files, \texttt{Scanner.java} and \texttt{Reader.java}. Each file has one conflict block. The first conflict block in \texttt{Scanner.java} has 3 conflicting LOC while the second conflict block in \texttt{Reader.java} has 3 conflicting LOC (assuming we treat the whole body of the \texttt{Read} class as one line here for better visualization).


\subsection{Motivating Example}

To understand how refactorings complicate merge scenarios, consider the example inspired by multiple real conflicts in Figure~\ref{example}. In the left branch (Figure~\ref{left}), the developer renames class \texttt{Listen} to \texttt{Read} in \texttt{Reader.java} and extracts the \texttt{notNull} and \texttt{validate} calls from \texttt{addListener} to a new method, \texttt{validateObject}. In the right branch (Figure~\ref{right}), the other developer: (1) moves class \texttt{Listen} from being an outer class in \texttt{Reader.java} into an inner class of class \texttt{Reader} in the same file, (2) renames method \texttt{validateReader} to \texttt{validateObject}, (3) renames method \texttt{addReader} to \texttt{scanReader} in \texttt{Scanner.java}, and (4) changes the code inside \texttt{addListener}.

As shown in Figure~\ref{git}, \git reports a conflict in file \texttt{Reader.java} because the developers rename \texttt{Listen} on one branch and moves it into class \texttt{Reader} on the other. Although both branches change the same lines of code, a smart merge tool could automatically merge these changes by considering their semantics and simply renaming the moved class \texttt{Listen} to \texttt{Read}, as shown in the ``ideal'' merge result in Figure~\ref{refMerge}. We refer to the \git conflict in \texttt{Reader.java} from Figure~\ref{git} as a \textit{false positive}, because it is a conflict that can be automatically resolved.
If the conflict cannot be automatically resolved and required manual intervention, we would refer to it as a \textit{true positive}.

\git reports another conflict in file \texttt{Scanner.java}, where the developer on the left branch extracts code from the same region that the right branch edits the code within \texttt{addListener}. To resolve this conflict, the developer needs to compare the code inside of the extracted method \texttt{validateObject} (which is not even highlighted as part of the conflict) with the conflicting code from the right parent shown in conflict block. Such a comparison is even worse if the method was extracted to a distant location in the file, or another file altogether. However, a merge tool that considers the semantics of extract method would realize that the changes from the right parent should be performed in the extracted method, rather than in \texttt{addListener} and that these changes can be merged, as shown in Figure~\ref{refMerge}.

Figure~\ref{refMerge} shows the ideal merge result for this scenario. This merge result avoids the unnecessary conflict in \texttt{Reader.java} by understanding the semantics of the rename and move operations. It also avoids the unnecessary conflict in \texttt{Scanner.java} by understanding the semantics of the extract method operation and applying the right branches changes in \texttt{validateObject}.
Note, however, that the ideal merge result also reports a conflict in \texttt{Reader.java} and \texttt{Scanner.java} for \texttt{validateObject}. 
By renaming \texttt{validateReader} to \texttt{validateObject} on the right branch and extracting a method with the same name on the left branch, the developers introduce an \textit{accidental override}, which could introduce bugs or critical errors that may not be discovered until their software is released.
\git fails to report this because the developers did not change the same lines of code. 
Such a case illustrates \git reporting a \textit{false negative}, where the merge tool should report a conflict because integrating these changes requires the developer's intervention, but instead \git silently merges the changes. 

\section{\ourtool: Refactoring-aware Operation-based Merging}




The high level idea of operation-based refactoring-aware merging is that if we invert refactorings before merging and then replay the refactorings, there will be no refactoring related conflicts to complicate the merge. Figure~\ref{overviewDiagram} presents an overview of our implementation of \ourtool, which consists of the following five steps. 

\begin{enumerate}[leftmargin=*]
	\item \textit{Detect and Simplify Refactorings}: We  use \textit{RefactoringMiner}, a state-of-the-art refactoring detection tool with 99.7\% precision and 94.2\% recall \cite{RefMiner} to detect refactorings in each commit between the base commit and each parent respectively. We check if each detected refactoring can be simplified and simplify the refactorings accordingly.
	\item \textit{Invert Refactorings}: We use the corresponding refactoring list from Step 1 to invert each refactoring until all covered refactorings have been inverted. 
	\item \textit{Merge}: We use \git to merge the left and right parents, $P_L'$ and $P_R'$, after all their refactorings have been inverted.
	\item \textit{Detect Refactoring Conflicts}: We compare the left and right refactoring lists for potential refactoring conflicts and commutative relationships and merge them into one list.
	\item \textit{Replay Refactorings}: We finally use the merged refactoring list to replay all non-conflicting refactorings.
\end{enumerate}

In this section, we focus on our implementation of the operation-based approach to enable it to work on top of \git, which makes some of the details different from MolhadoRef.

\subsection{Step 1: Detect and Simplify Refactorings}


\paragraph*{\textit{Refactoring Detection}} We use \texttt{RefactoringMiner} to detect refactorings in each commit between the base commit and each parent commit respectively. 
We detect refactorings in each commit instead of only comparing the base and parent commits to ensure precise detection in longer histories. This is an important difference from \texttt{MolhadoRef} as the use of RefactoringMiner allows \ourtool to be implemented for Git, instead of relying on a research-based VCS.

\paragraph*{\textit{Refactoring Simplification}} \ourtool processes each detected refactoring one by one and keeps a list of processed refactorings, \textit{ProcessedRefList}, for the left and right branches. 
We compare each detected refactoring to the refactorings in \textit{ProcessedRefList} to determine if it is either a transitive refactoring or part of a refactoring chain (defined below). 

We define \textit{transitive refactorings} as successive related refactorings of the same refactoring type.  For example, consider that method \texttt{foo} is renamed to \texttt{bar}. In the next commit, \texttt{bar} is renamed to \texttt{foobar}. In this case, the two method renamings are transitive and \texttt{foo} is eventually being renamed to \texttt{foobar}.
When \ourtool finds that a newly detected refactoring is a transitive refactoring of an existing refactoring in \textit{ProcessedRefList}, it updates the related transitive refactorings in \textit{ProcessedRefList} instead of adding a new refactoring. In this example, \ourtool would first add \textit{rename \texttt{foo} to \texttt{bar}} to \textit{ProcessedRefList}. When it processes \textit{rename \texttt{bar} to \texttt{foobar}}, it detects that this is a transitive refactoring of an existing refactoring in \textit{ProcessedRefList} so it will simply update the existing refactoring to \textit{rename \texttt{foo} to \texttt{foobar}}.

\textit{Refactoring chains} consist of two or more refactorings that touch the same program element. When two refactorings touch the same program element, the details of that program element will diverge from what is stored in RefactoringMiner's refactoring object, causing the refactoring to not be found when later inverting the refactoring or detecting refactoring conflicts. For example, when a method is renamed in \textit{class A} and \textit{class A} is renamed to \textit{class B} in a later commit, the first refactoring object will still associate the method with \textit{class A}. This means that if a transitive refactoring is later performed on the same method, we will not be able to detect the transitive relationship because the methods will be associated with different classes.

Therefore, when we find that a refactoring is part of a refactoring chain, we update the refactorings in the refactoring chain. 
For example, consider that after \textit{A.foo} is renamed to \textit{A.bar}, \textit{class A} is renamed to \textit{class B}. Then in a later commit, \textit{B.bar} is renamed to \textit{B.foobar}. Since method \textit{foo} was renamed to \textit{bar} inside of class \textit{A}, \textit{A.bar} and \textit{B.bar} have different method signatures and the information that these \textit{Rename Method}s are transitive is lost. 
To address this, \ourtool first adds the first refactoring \textit{rename A.foo to A.bar} to the list.
When it later processes the second refactoring \textit{rename class A to B}, it adds this second refactoring to the list \textit{and} also updates the first refactoring to \textit{rename B.foo to B.bar}.
That way, when \ourtool processes the third refactoring \textit{rename B.bar to B.foobar}, it can detect the transitive relationship and update it accordingly.
Our artifact contains all our detailed logic for detecting transitive refactorings and refactoring chains~\cite{artifact}.

\paragraph*{\textit{Refactoring Order}} Since we do not know the order in which developers performed refactorings within the same commit, we cannot simply invert the refactorings in the opposite order they are detected in. Instead, we reorder the refactorings in a top-down order based on the granularity of the program element being refactored. For example, class level refactorings come before method level refactorings. 

Combining transitive refactorings, updating refactoring chains and using a top-down order has three advantages. First, when inverting and replaying refactorings, all transitive refactorings are combined and can be treated as if they were detected at a coarse-grained granularity. 
This is an important distinction from MolhadoRef, because it reduces the number of refactorings that need to be performed and simplifies conflict detection while at the same time ensuring precise refactoring detection. 
Second, the combination of updating refactoring chains and using the pre-determined order removes any need to keep track of the order that the refactorings are detected in. 
Lastly, using a top-down order while simplifying refactorings automatically breaks any circular dependencies between refactoring operations. This is another important difference from MolhadoRef which required user intervention to help resolve circular dependencies; by automatically breaking circular dependencies, \ourtool allows the user to focus on the conflicts. 

\subsection{Step 2: Invert Refactorings}

\begin{table}[t!]
    \centering
         \caption{\revision{The list of 17 refactorings supported by \ourtool and their corresponding inverse refactorings.}}
    \label{tab:inverse-refactorings}
    \resizebox{0.5\textwidth}{!}{
    \begin{tabular}{ll}
    \toprule
     \textbf{Refactoring}&\textbf{Inverse Refactoring}  \\
     \midrule
     \texttt{RenameMethod(m$_1$, m$_2$)}&\texttt{RenameMethod(m$_2$, m$_1$)}\\
     \texttt{MoveMethod(c$_1$, c$_2$)}&\texttt{MoveMethod(c$_2$, c$_1$)}\\
     \texttt{Move\&RenameMethod(c$_1$.m$_1$, c$_2$.m$_2$)}&\texttt{Move\&RenameMethod(c$_2$.m$_2$, c$_1$.m$_1$)}\\
     \texttt{RenameClass(c$_1$, c$_2$)}&\texttt{RenameClass(c$_2$, c$_1$)}\\
     \texttt{MoveClass(loc$_1$, loc$_2$)}&\texttt{MoveClass(loc$_2$, loc$_1$)}\\
     \texttt{Move\&RenameClass(loc$_1$.c$_1$, loc$_2$.c$_2$)}&\texttt{Move\&RenameClass(loc$_2$.c$_2$, loc$_1$.c$_1$)}\\
     \texttt{InlineMethod(m$_1$, m$_2$.l$_x$-l$_y$)}&\texttt{ExtractMethod(m$_2$.l$_x$-l$_y$, m$_1$)}\\
    \texttt{ExtractMethod(m$_1$.l$_x$-l$_y$, m$_2$)}&\texttt{InlineMethod(m$_2$, m$_1$.l$_x$-l$_y$)}\\
    \texttt{PullUpMethod(c$_x$-c$_y$,c$_2$)}&\texttt{PushDownMethod(c$_2$,c$_x$-c$_y$)}\\
    \texttt{PushDownMethod(c$_1$,c$_x$-c$_y$)}&\texttt{PullUpMethod(c$_x$-c$_y$,c$_1$)}\\
    \texttt{RenameField(f$_1$, f$_2$)}&\texttt{RenameField(f$_2$, f$_1$)}\\
    \texttt{MoveField(c$_1$, c$_2$)}&\texttt{MoveField(c$_2$, c$_1$)}\\
    \texttt{Move\&RenameField(c$_1$.f$_1$, c$_2$.f$_2$)}&\texttt{Move\&RenameField(c$_2$.f$_2$, c$_1$.f$_1$)}\\
    \texttt{PullUpField(c$_x$-c$_y$,c$_2$)}&\texttt{PushDownField(c$_2$,c$_x$-c$_y$)}\\
    \texttt{PushDownField(c$_1$,c$_x$-c$_y$)}&\texttt{PullUpField(c$_x$-c$_y$,c$_1$)}\\
    \texttt{RenamePackage(p$_1$, p$_2$)}&\texttt{RenamePackage(p$_2$, p$_1$)}\\
    \texttt{RenameParameter(p$_1$, p$_2$)}&\texttt{RenameParameter(p$_2$, p$_1$)}\\
    \bottomrule
    \end{tabular}
}
\end{table}

Once refactorings are detected, \ourtool creates a refactoring-free version of each parent commit by inverting the refactorings in the \textit{ProcessedRefList} on each branch from Step 1. 
To invert a refactoring $r$, \ourtool needs to create and apply the inverse refactoring $\bar{r}$.  $\bar{r}$ is an inverse of $r$ if $\bar{r}(r(E)) = E$. For example, the inverse of a refactoring that renames method \texttt{foo} to \texttt{bar} is another refactoring that renames \texttt{bar} to \texttt{foo}.
\revision{Table~\ref{tab:inverse-refactorings} provides a list of each refactoring and the refactoring operation that \ourtool uses to invert it.}

\ourtool uses the information provided by \textit{RefactoringMiner} to create each inverse refactoring. Each refactoring detected by \textit{RefactoringMiner} is represented by a data structure that contains important information about the refactoring. Among others, the data structure contains information such as the refactoring type, information about the original program element, and information about the refactored program element. From the provided information, \ourtool obtains the corresponding elements and executes the refactoring through a refactoring engine. Importantly, executing the inverse refactoring does not only invert the refactored program element, but it also changes any references to the program element. This includes references added at any point after the refactoring was performed. In the case that the refactored program element is deleted in a future commit, the inverse refactoring cannot be performed and \ourtool moves on to the next refactoring. 

\subsection{Step 3: Merge}

After all refactorings are inverted on both branches, only non-refactoring changes remain in the parent commits. We refer to this version of each parent as $P'$ in Figure~\ref{overviewDiagram}. In this step, we textually merge $P_L'$ and $P_R'$.  Most same-line or same-block conflicts that would have been caused by refactorings are now eliminated through inverting the refactorings. However, some same-line and same-block conflicts may still exist because additional edits may have been performed to or beyond the refactored code. 

For example, consider the conflict blocks in \texttt{Scanner.addListener} in Figure~\ref{example}. If the developer adds several other lines of code to the extracted method, those lines will be inlined to the \texttt{validateObject} method invocation and reported in the conflict block. In this case, \ourtool will report more conflicting lines than \git because no matter how many lines of code are added to \textit{Scanner.validateObject}, Git's conflicting region will remain the same. While the extra conflicting lines that \ourtool reports could be considered to be disadvantageous, inlining the extracted code clearly indicates what code is part of the conflict in a single location.

\subsection{Step 4: Detect Refactoring Conflicts}

Generally speaking, a pair of refactorings that touch unrelated program elements do not have any interaction. However, a pair of refactorings that touch related program elements will have interactions, which can be conflicting or commutative. For each pair of refactorings, we have to predetermine the interactions that the refactorings can result in and then use that knowledge to detect conflicts and commutative refactorings. Refactoring operations that \textit{conflict} cannot both be replayed, while refactoring operations that are \textit{commutative} can be replayed in either order and will result in the same code. We make the assumption that two refactoring operations cannot both conflict and have a commutative relationship. We carefully compute and revise the conflict and commutative logic for each refactoring combination, which we explain below and can be found in our artifact~\cite{artifact}. \ourtool uses this knowledge to compare each refactoring in the left branch with each refactoring in the right branch and detect refactoring conflicts.

\subsubsection{Detecting Conflicts}

\ourtool first checks if the two refactoring operations are conflicting. There are a series of preconditions that must be met for two refactoring operations to conflict.
To illustrate, we provide an example using the conflict logic for $RenameMethod(m_1,m_2)$ and $RenameMethod(m_3,m_4)$ \revision{in Equation~\ref{eq:conflict}.

\begin{equation}
\small
    \begin{aligned}
hasConflict(RenameMethod(c_1.m_1,c_2.m_2),\\RenameMethod(c_3.m_3,c_4.m_4)) := \\
((c_1.m_1 == c_3.m_3 \land c_2.m_2 \neq c_4.m_4) \\\lor (c_1.m_1 \neq c_3.m_3 \land c_2.m_2 == c_4.m_4))\\  \lor 
(\neg overrides(c_1.m_1,c_3.m_3) \land overrides(c_2.m_2,c_4.m_4)\\ \lor 
(\neg overloads(c_1.m_1,c_3.m_3) \land overloads(c_2.m_2,c_4.m_4) \\ 
    \end{aligned}
\label{eq:conflict}
\end{equation}
}
These two refactorings result in a conflict if (1) the source of both refactorings is the same program element ($c_1.m_1 = c_3.m_3$) but their destinations differ ($c_2.m_2 \neq c_4.m_4$) or (2) the sources of both renames are different program elements ($c_1.m_1 \neq c_3.m_3$) but the renamed destinations are the same program element ($c_2.m_2 = c_4.m_4$). In other words, if the same method is renamed to two separate names or if two different methods inside of the same class are renamed to the same name with the same signature, then the refactorings conflict. 

In addition, two refactoring operations can conflict without changing the same program element. We refer to this as a \textit{semantic conflict}. 
There are two examples of semantic conflicts for $RenameMethod$/$RenameMethod$: (1) an accidental overload and (2) an accidental override. In the case of an accidental overload, two methods with different names are renamed to the same name in the same class but have different signatures. 
In the case of an accidental override, two methods within classes with an inheritance relationship are renamed to the same name with the same signature, which causes one of the methods to override the other. 
Semantic conflicts will not be detected by a text-based merge tool such as Git because the same line is not changed by both branches. The developer might not realize the problem until it appears in testing, or worse in production. The \textit{Rename Method} and \textit{Extract Method} refactorings are an example of conflicting refactorings that can cause an accidental override, such as the motivating example in Figure~\ref{example}. 


\subsubsection{Detecting Commutative Relationships}

After RefMerge checks for refactoring conflicts, it checks for a \textit{commutative relationship} between the two refactoring operations using the corresponding predetermined commutative logic. Two refactoring operations can only be commutative if they do not conflict and if they are different types of refactorings. If the pair of refactorings meets these conditions and they both refactor the same program element, then they are commutative. For example, \textit{Rename Method} and \textit{Rename Method} cannot be commutative because they are the same refactoring type and there is no way the same program element can be renamed on both branches to different names without conflicting. However, \textit{Move Method} and \textit{Rename Method} can be performed on the same program element commutatively. Similarly, the \textit{Move Class} and \textit{Rename Class} refactorings performed on class \textit{Listen} in Figure~\ref{example} are an example of commutative refactorings. 


\revision{We present the commutative logic for $MoveMethod(m_1,m_2)$ and $RenameMethod(m_3,m_4)$
as an example in Equation~\ref{eq:comm}.

\begin{equation}
\small
    \begin{aligned}
       isCommutative(MoveMethod(c_1.m_1,c_2.m_2),\\RenameMethod(c_3.m_3,c_4.m_4) := \\
(c_1.m_1 == c_3.m_3 \land c_2.m_2 \neq c_4.m_4) 
    \end{aligned}
\label{eq:comm}
\end{equation}
}
These two refactorings are commutative if the source of both refactorings is the same program element ($m_1 = m_3$) and their destinations are different ($m_2 \neq m_4$). The idea is that if a \textit{Move Method} and \textit{Rename Method} refactoring are performed on the same program element, then we can move the program element and then rename it, or rename it and then move it.

After all detected refactorings have been compared between branches for refactoring conflicts and commutative relationships, \ourtool combines the refactoring lists containing non-conflicting refactorings from each branch into one list. 
 While \ourtool inverts the refactorings on each branch in a top-down order (after simplifying the refactoring lists to enable this), it orders the combined refactoring list in a bottom-up order for replaying refactorings.
 Multiple refactorings might touch the same program element, such as a \textit{Move Method} and a \textit{Rename Class}. By renaming the class before moving the method, \ourtool will not be able to find the method refactoring, because the class that the method is moved from will no longer exist. 
 Since higher-level program elements do not depend on lower level program elements, replaying refactorings bottom-up allows \ourtool to replay the refactorings without any additional effort. The replay refactoring list for Figure~\ref{example} after detecting refactoring conflicts and commutative
conflicts would contain \textit{Rename Method} \texttt{addReader} to \texttt{scanReader} and  \textit{Move And Rename Class} \texttt{Listen} to inner class \texttt{Reader.Read}. The conflicting refactoring list would contain \textit{Extract Method} \texttt{validateObject} from \texttt{addListener} and \textit{Rename Method} \texttt{validateReader} to \texttt{validateObject}.

\subsection{Step 5: Replay Refactorings}

Finally, \ourtool replays the refactorings. For each inverted refactoring, \ourtool re-creates and performs the refactoring that was originally performed by the developer. Executing the refactoring includes updating all references in the program, including those added on the other branch.

\subsection{Current Implementation}

\paragraph*{\textit{Technologies and Tools}} We implement \ourtool as an IntelliJ\footnote{https://www.jetbrains.com/idea} plugin for merging Java programs. It consists of four key modules 
corresponding to the steps of the proposed technique. We use 
\textit{RefactoringMiner} \cite{RefMiner} to detect the refactorings and use the \textit{IntelliJ refactoring engine} to programatically invert and replay the refactorings.

\paragraph*{\textit{Supported Refactorings}} Even though the idea of operation-based refactoring-aware merging and our proposed implementation of it generally applies to all refactorings, there are more than 70 known refactoring types \cite{Fowler}; it is a large engineering effort to implement every refactoring. Instead of implementing every refactoring, we use a subset of 17 refactorings to show the feasibility of the approach and enable the empirical comparison. 


\revision{
Note that \intellimerge supports 21 refactorings, including the 15 refactorings supported by Mahmoudi et al.'s analysis~\cite{SANER}. 
While the \intellimerge paper does not explicitly state which refactorings \intellimerge supports, 13 of the refactorings can be inferred from the matching rules in the publication~\cite{IntelliMerge}.
However, we find 8 more refactorings included in the tool's implementation.
When deciding what refactorings \ourtool supports, we prioritized the 13 refactoring types inferred from the paper.
In addition, we added support for \textit{Rename And Move Field}, \textit{Method}, and \textit{Class} to get more complete coverage at various program element granularities. 
Finally, we added support for \textit{Rename Parameter} to add some coverage at the parameter level.
Overall, \ourtool supports a subset of 17 out of the 21 refactorings \intellimerge supports.

}

When a refactoring is performed that \ourtool does not support or \ourtool fails to invert, \ourtool results in the same merge as \git for the program element. Thus, \ourtool should improve on \git for supported refactorings, but should be no worse than \git for refactorings that are not currently supported. 
It is worth noting that our open-source implementation of \ourtool is designed in a modular way to easily allow for extension. 
\revision{In general, adding a new refactoring requires adding a handler to invert and replay the refactoring and adding logic for how it interacts with the existing refactoring. Our artifact~\cite{artifact} contains a step-by-step guide for adding a new refactoring}.


\section{Evaluation Setup}

We compare the effectiveness of \ourtool, \git, and the state-of-the-art refactoring-aware merge tool, \intellimerge \cite{IntelliMerge} on \sample merge scenarios that contain refactoring-related conflicts from 20 open-source projects.
These projects include the original 10 projects \intellimerge was evaluated on as well as an additional 10 projects with different distributions of conflicting merge scenarios.
We answer the following research questions:

\begin{enumerate}[label=\textbf{RQ\arabic*}, leftmargin=*]
    \item \textit{How many merge conflicts do the three merge tools report?} A tool that automatically resolves more merge conflicts will reduce the time and effort developers have to spend resolving conflicts. We report conflicts at all granularity levels (scenarios, files, and conflict blocks). 

    \item \textit{What are the discrepancies between the merge conflicts that RefMerge and IntelliMerge report?} While either tools may report less conflicts, which seems better at face value, we need to investigate if they correctly resolve the conflicts or if they miss reporting real conflicts. We perform a qualitative analysis on the results reported by \ourtool and \intellimerge to understand the strengths and weaknesses of each tool.
\end{enumerate}


\subsection{Project \& Merge Scenario Selection}

\begin{table*}
\centering
\caption{\revision{Number of conflicting merge scenarios with involved refactorings for the 20 projects we evaluate on. The 10 projects from the \intellimerge paper are in bold. Our evaluation is based on the merge scenarios in Column c.}}
\label{tab:conflict-scenarios}
\resizebox{0.98\textwidth}{!}{
\begin{tabular}
{lrr>{\raggedleft\arraybackslash}p{2.75cm}
>{\raggedleft\arraybackslash}p{3.25cm}
>{\raggedleft\arraybackslash}p{4cm} 
>{\raggedleft\arraybackslash}p{4.2cm}} 
\toprule
\textbf{Project}&\textbf{Stargazers}&\textbf{Merge Scenarios}&
{\textbf{a. Conflicting Merge Scenarios}}
& {\textbf{b. Conflicting Java Merge Scenarios (\% from Col. a)}}
&\textbf{c. Conflicting Java Merge Scenarios w/ Involved Refactorings (\% from Col. b)}
&\textbf{d. Conflicting Java Merge Scenarios w/ \textit{Only} Involved Refactorings (\% from Col. c)}\\
\midrule
\textbf{cassandra}            & 6,882 & 10,719 & 4,509 (42\%) & 2,693 (25\%)  & 922 (34\%) & 244 (27\%)   \\
\textbf{elasticsearch}        & 56,665 & 5,111  & 561 (11\%) & 504 (9\%) & 178 (35\%) & 30  (17\%)  \\
\textbf{gradle}               & 12,410 & 7,690 & 1,127 (15\%) & 300 (3\%) & 117 (39\%) & 28 (24\%)   \\
\textbf{antlr4}               & 10,738 & 1,935 & 398 (21\%) & 198 (10\%) & 100 (51\%) & 14 (14\%)   \\
platform\_frameworks\_support & 1,609  & 67,584 & 3,690 (55\%) & 570 (15\%) & 96 (17\%) & 25 (26\%)    \\
\textbf{deeplearning4j}       & 12,208 & 6,997 & 566 (8\%) & 386 (6\%) & 93 (24\%) & 20 (22\%)   \\
\textbf{realm-java}           & 11,206 & 3,405 & 683 (20\%) & 312 (9\%) & 92 (29\%) & 23 (25\%)    \\
jackson-core                  & 1,984  & 584 & 318 (54\%) & 201 (34\%) & 81 (40\%) & 17  (21\%)  \\
android                       & 3,161  & 1,805 & 315 (17\%) & 208 (12\%) & 81 (39\%) & 9 (11\%)    \\
cometd                        & 535    & 759 & 369 (49\%) & 173 (23\%) & 63 (36\%) & 11 (18\%)   \\
\textbf{storm}                & 6,278  & 3,626 & 267 (7\%) & 87 (2\%) & 33 (38\%) & 12 (36\%)    \\
ProjectE                      & 308    & 386 & 79 (20\%) & 73 (19\%) & 30 (41\%) & 3 (10\%)  \\
\textbf{javaparser}           & 3,859  & 2,427 & 94 (4\%) & 73 (3\%) & 23 (32\%) & 6 (26\%)     \\
druid                         & 24,576 & 1,498 & 151 (10\%) & 138 (9\%) & 17 (12\%) & 3 (18\%)    \\
androidannotations            & 11,171 & 739 & 73 (10\%)&  71 (10\%) & 15 (21\%) & 1 (7\%)  \\
\textbf{junit4}               & 8,198  & 400 & 46 (11\%) & 41 (10\%) & 14 (34\%) & 2 (14\%)    \\
MinecraftForge                & 4,945  & 792 & 77 (10\%) & 46 (6\%) & 14 (30\%) & 3 (21\%)    \\
iFixitAndroid                 & 143    & 171 & 29 (17\%) & 24 (14\%) & 13 (54\%) & 0 (0\%)   \\
MozStumbler                   & 609    & 934 & 32 (3\%) & 26 (3\%) & 10 (38\%)  & 1 (10\%)  \\
\textbf{error-prone}          & 5,717  & 133 &  24 (18\%) & 21 (16\%) & 9 (43\%) & 1 (11\%) \\ 
\midrule
Total                         &        & && & 2,001 & 453  \\
\bottomrule
\end{tabular}
}
\end{table*}

We first include the same 10 projects that the \intellimerge authors use in their evaluation \cite{IntelliMerge}. To select these projects, the authors searched for the top 100 Java projects with high numbers of stargazers on Github, and then selected the projects with the most merge commits and contributors \cite{IntelliMerge}. 
The authors then ran the analysis by Mahmoudi et al.~\cite{SANER} on these 10 projects to identify conflicting merge scenarios that have refactoring changes involved in the conflict.
In a nutshell, this analysis replays merge scenarios in the \git history to find conflicting ones, uses RefactoringMiner~\cite{RefMiner} to find refactorings in the history of these conflicting merge scenarios, and then compares the location of the refactorings to the location of the conflict blocks to determine if a conflict has an \textit{involved refactoring}.
At the time of the \intellimerge publication, these 10 projects contained 1,070 conflicting merge scenarios with involved refactorings.

For generalizability, we expand our evaluation to cover an additional 10 projects. 
Mahmoudi et al.~\cite{SANER} shared a data set with the results of their analysis for 2,955 open-source GitHub projects.
We use this data set to select the additional 10 projects for our evaluation.
Our goal is to have a selection of projects with different distributions of (conflicting) merge scenarios to avoid any bias towards project-specific practices.
Thus, we sort the 2,955 projects within the dataset based on the number of refactoring-related conflicts each project has. We randomly select three projects from the bottom 30\% of the projects, four from the middle 40\%, and three from the top 30\%.

Given the 20 selected projects, we collect an up-to-date set of merge scenarios with involved refactorings by re-running Mahmoudi et al.'s analysis~\cite{SANER} on the latest history of each project as of September 26, 2021.
Our artifact page~\cite{artifact} contains the exact version of each project that we consider.
This means that for the 10 projects originally used by \intellimerge, our data set contains the original 1,070 merge scenarios as well as any additional ones that appear in the \git history since their publication date.

\begin{figure*}[t]
\centering
\includegraphics[width=0.9\textwidth]{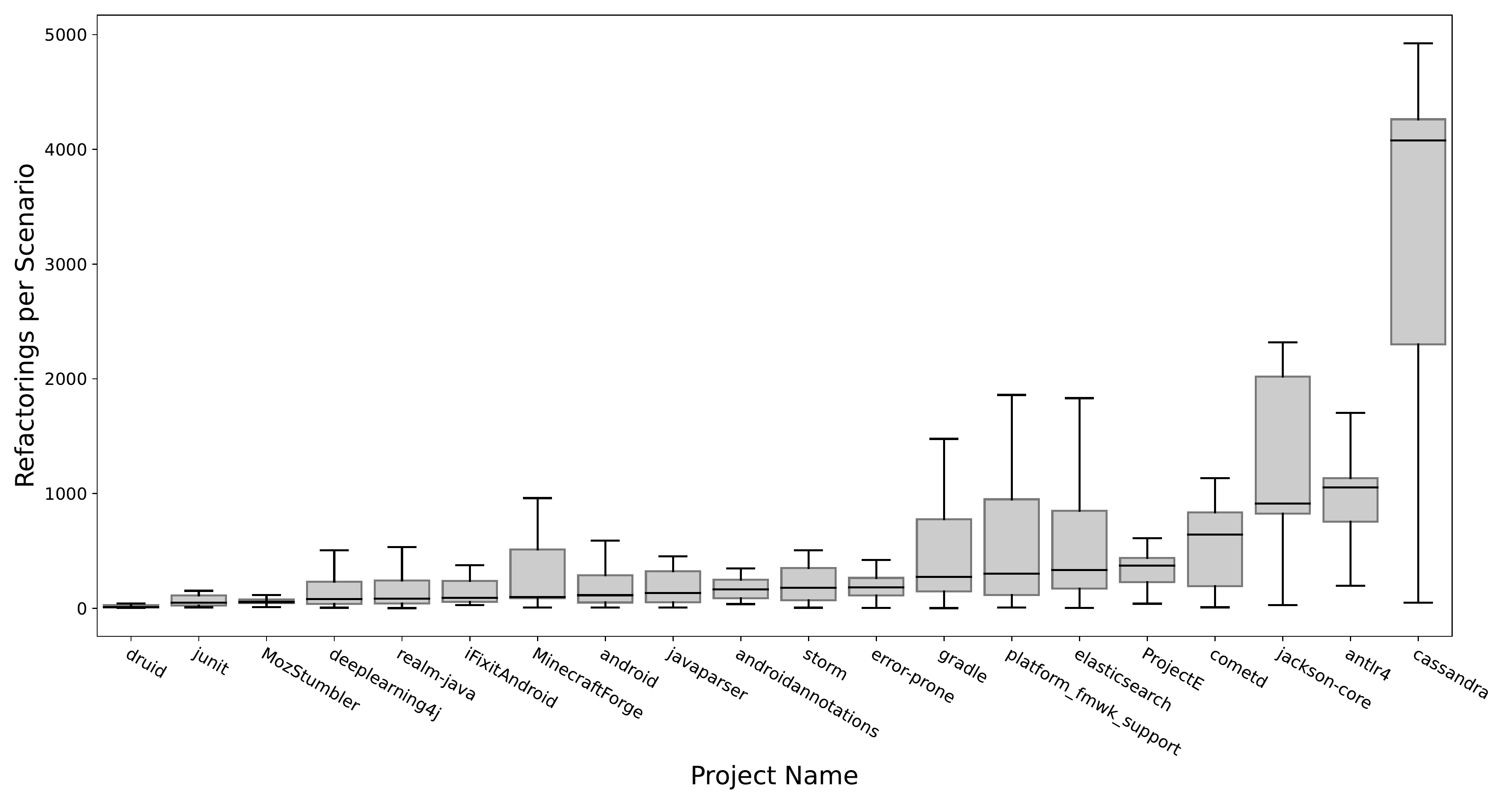}
\caption{\revision{Distribution of refactorings per merge
scenario that we evaluate on in each project}
}
\label{fig:refactoring-distribution}
\vspace{-4mm}
\end{figure*}

Table~\ref{tab:conflict-scenarios} shows the number of merge scenarios in each project, the number of conflicting merge scenarios, the number of conflicting merge scenario with conflicting Java files, the number of merge scenarios with refactoring-related Java conflicts, and the number of merge scenarios with \textit{only} refactoring-related Java conflicts. 
The table shows that the frequency of merging and frequency of conflicting merges varies between projects.
Thus, our evaluation covers projects with frequent merge scenarios (e.g., cassandra and platform\_frameworks\_support) as well as those with infrequent merges in their history (e.g., error-prone or iFixitAndroid). We also cover both projects that are conflict prone (e.g., cassandra and jackson-core) as well as those with infrequent conflicts, regardless of frequency of merging (e.g., storm and javaparser).
Table~\ref{tab:conflict-scenarios} also shows the projects used in the IntelliMerge paper in bold and provide the number of stargazers of each project. 
\revision{
Overall, we evaluate on \checkNum{2,001} conflicting merge scenarios with involved refactorings.
Figure~\ref{fig:refactoring-distribution} shows the distribution of refactorings per merge scenario that we evaluate on in each project. 
As shown, the median number of refactorings per scenario varies widely among the projects, with \textit{druid} having the lowest median and \textit{cassandra} having the highest.
}



\subsection{Reproducing \intellimerge}
Before describing the evaluation metrics we use for comparing the merge tools, we need to ensure that we are correctly running \intellimerge.
Thus, we first attempt to reproduce the results found in the corresponding publication~\cite{IntelliMerge} using their exact setup and data, as shared in their Github repository~\cite{IntelliMergeURL}. We share the exact steps we followed as well as the details of the results of reproducing \intellimerge\footnote{\label{replication}https://github.com/max-ellis/IntelliMerge/tree/evaluation}.

We run \intellimerge v1.0.7 on the same 1,070 merge scenarios used in the original publication, including their same post-processing steps such as removing all comments from the merged files.
We use the same calculation proposed by \intellimerge's authors to measure precision and recall for \intellimerge and Git. They propose comparing the auto-merged code with the manually-merged code to measure precision and recall. They define \textit{auto-merged code} as code that is not part of a conflict block in a tool's merge result and \textit{manually-merged code} as the code that appears in the resolved merge commit in the git history. We use the same \textit{diff} tool provided by \git that the \intellimerge authors used to calculate the number of different lines between the auto-merged and manually-merged code.
Note that \intellimerge reports precision and recall based \textit{only} on the conflicting files in each merge scenario, not on all changed files in the scenario.

We were not able to reproduce the exact numbers found in the \intellimerge paper~\cite{IntelliMerge}. After emailing the authors, we verified that they perform manual post-processing steps to deal with some cases that are caused by the program elements being in a different order as well as format related diffs, such 
as textually moving, reordering, and cosmetic diffs. Because of these undocumented manual post-processing steps, it is impossible to reproduce the exact numbers in the \intellimerge paper. Although we were not able to get the exact numbers, the precision and recall we obtained were within 10\% of the numbers in their paper. For further confirmation, we explicitly shared our setup\footnotemark[\getrefnumber{replication}] with the \intellimerge authors and received confirmation that our setup is correct and that the differences in results we obtained do not misrepresent \intellimerge.



\subsection{Tool Comparison Setup}

After verifying with the \intellimerge authors that we are correctly running their tool, we could proceed with our evaluation.
Given the \checkNum{2,001} merge scenarios, we identify the base commit, left parent commit, and right parent commit of each scenario.
We provide each tool (\git, \intellimerge, \ourtool) with these three commits in order to perform its three-way merge.
We record the results of \textit{all changed files} in the merge scenario, as opposed to only conflicting files (which is what the \intellimerge evaluation does).
Considering the result of all changed files allows us to catch cases where one of the tools introduces a conflict in a file that \git did not originally report a conflict for.
Additionally, while the \intellimerge authors removed comments in their evaluation, we do not post-process the results of any of the merge tools in any way to ensure that we see the same results a developer using the tool in practice would see. 
Overall, our goal in this evaluation is to minimize any manual pre and post processing steps such that we can compare the results of these tools in a practical setting.

Note that while the scenarios we evaluate on may have refactorings that either tools do not support, we do not limit the evaluation to only supported refactorings so we can also understand how the tools handle unsupported refactorings.

We run our experiments on a quad-core computer with \revision{Intel (R) Core (TM) i5-12600K CPU @ 3.70GHz, 32 GB RAM} and Ubuntu 20.04 OS. For feasibility of completing the evaluation, we use a 15 min timeout for each tool.

\subsection{Used Metrics and Analysis Methods} 

We choose not to use the same recall and precision metrics that the \intellimerge authors propose, because (1) these metrics do not correctly capture the effectiveness of a merge tool and (2) auto-merged code is not a reliable way to measure false positives and false negatives. 

Consider the merge conflict in \texttt{Scanner.java} in Figure~\ref{git}. If the developer chose to merge the changes from the left branch in the manual merge, the manually merged code will have 15 lines of code. Meanwhile, if a merge tool always accepts changes from both branches, then the auto-merged code will have 17 lines of code.  In this case, \git diff will report 2 different lines since the auto-merged code also contains changes from the right branch while the manually merged code does not. In this case, the recall will be 1 and while the precision will not be 1, it will still be high (88\%) and will not reflect the fact that the tool failed to detect the conflict.
In their threats, the \intellimerge authors themselves recognize that using manually committed code as the ground truth is unreliable, because manually committed files often contain mistakes. 

Instead, in RQ1, we report the number of conflicts each tool detects at various granularity levels (scenarios, files, and conflict regions).
Additionally, we do not only report these numbers in isolation but instead report them at a scenario level to understand the proportion of scenarios in which each tool can improve the situation for a developer.
Additionally, for RQ2, we manually sample merge conflicts that differ between the merge tools to understand the quality of the merge results and how the behavior of these tools differ in handling different types of merge scenarios. A similar analysis has been used in the past by Cavalcanti et al.~\cite{Cavalcanti} to get a better understanding of merge results.

\newcolumntype{M}[1]{>{\centering\arraybackslash}m{#1}}

\section{RQ1: Quantitative Tool Comparison}

\begin{table*}[t]
\centering
\caption{Merging results for each tool, compared to Git. Number in parentheses shows the proportion from total scenarios in each project. For each project, the tool that was able to completely resolve more merge scenarios is shown in bold.}
\label{tab:scenario-breakdown}
\resizebox{0.95\textwidth}{!}{
\begin{tabular}{@{}lrrrrrrrrr@{}}
\toprule
\multirow{2}{*}{\textbf{Project Name}} & \multirow{2}{*}{\textbf{Total Scenarios}} & \multicolumn{4}{c}{\textbf{IntelliMerge}}                                    & \multicolumn{4}{c}{\textbf{RefMerge}}                                        \\ \cmidrule(lr){3-6} \cmidrule(lr){7-10}
&                                           & \textbf{Resolved} & \textbf{Changed} & \textbf{Unchanged} & \textbf{Timeout} & \textbf{Resolved} & \textbf{Changed} & \textbf{Unchanged} & \textbf{Timeout} \\ 
\midrule
\revision{cassandra}                              & 922                                       & 33 (4\%)         & 108 (12\%)         & 1 (0\%)            & 780 (85\%)       & \textbf{54 (6\%) }         & 204 (22\%)         & 322 (35\%)         & 342 (37\%)       \\
\revision{elasticsearch}                          & 178                                       & 3 (2\%)           & 103 (58\%)        & 0 (0\%)            & 71 (40\%)        & \textbf{9 (5\%)}           & 45 (25\%)        & 85 (48\%)          & 39 (22\%)        \\

\revision{gradle}                   & 118                                       & 1 (1\%)               & 106 (90\%)    & 0 (0\%)   & 11 (9\%)             & \textbf{9 (8\%)}     & 33 (28\%)     & 75 (64\%) & 1 (1\%)   \\
\revision{antlr4}                   & 100                                       & 1 (1\%)               & 96 (96\%)     & 0 (0\%)   & 3 (3\%)              & \textbf{4 (4\%)}      & 39 (39\%)     & 57 (57\%) & 0 (0\%)   \\
\revision{platform\_fwk\_support}   & 95                                        & 5 (5\%)               & 56 (59\%)     & 1 (1\%)   & 33 (35\%)            & \textbf{9 (10\%)}        & 40 (42\%)   & 46 (48\%) & 0 (0\%)   \\
\revision{deeplearning4j}           & 93                                        & 3 (3\%)               & 89 (96\%)     & 0 (0\%)   & 1 (1\%)               & \textbf{5 (5\%)}       & 31 (33\%)   & 57 (61\%) & 0 (0\%)   \\ 
\revision{realm-java}               & 92                                        & \textbf{7 (8\%)}               & 82 (89\%)     & 1 (1\%)   & 2 (2\%)               & \textbf{7 (8\%)}      & 32 (35\%)    & 53 (58\%) & 0 (0\%)   \\
\revision{jackson-core}             & 81                                        & 0 (0\%)               & 81 (100\%)    & 0 (0\%)   & 0 (0\%)               & \textbf{3 (4\%)}      & 25 (31\%)     & 52 (64\%) & 0 (0\%)   \\
\revision{android}                  & 81                                        & 3 (4\%)               & 78 (96\%)     & 0 (0\%)   & 0 (0\%)               & \textbf{8 (10\%)}     & 28 (35\%)     & 45 (56\%) & 0 (0\%)   \\
\revision{cometd}                   & 63                                        & 2 (3\%)               & 59 (93\%)     & 1 (2\%)   & 1 (2\%)               & \textbf{4 (6\%)}      & 20 (32\%)     & 39 (62\%) & 0 (0\%)   \\
\revision{storm}                    & 33                                        & \textbf{1 (3\%)}              & 30 (91\%)     & 1 (3\%)   & 1 (3\%)         & \textbf{1 (3\%)}    & 13 (39\%)     & 19 (58\%) & 0 (0\%)   \\
\revision{ProjectE}                 & 30                                        & \textbf{1 (3\%)}              & 28 (94\%)     & 0 (0\%)    & 1 (3\%)         & 0 (0\%)       & 13 (43 \%)      & 17 (57\%)         & 0 (0\%)   \\
\revision{javaparser}               & 23                                        & \textbf{3 (13\%)}     & 19 (83\%)     & 1 (4\%)       & 0 (0\%)         & 2 (9\%)           & 9 (39\%)          & 12 (52\%)         & 0 (0\%)       \\
\revision{druid}                    & 17                                          & \textbf{3 (18\%)}           & 15 (89\%)      & 0 (0\%)               & 0 (0\%)       & 2 (12\%)       & 9 (53\%)      & 6 (35\%)       & 0 (0\%)   \\
\revision{androidannotations}        & 15                                         & \textbf{1 (7\%)}            & 14 (93\%)      & 0 (0\%)              & 0 (0\%)         & 0 (0\%)                  & 9 (60\%)         & 6 (40\%)   & 0 (0\%)       \\
\revision{junit4}                     & 14                                        & \textbf{1 (7\%)}                     & 13 (93\%)       &  0 (0\%)             & 0 (0\%)        & \textbf{1 (7\%)}           & 8 (57\%)               & 4 (29\%) & 0 (0\%)   \\
\revision{MinecraftForge}             & 14                                        & 1 (7\%)                     & 10 (72\%)       & 0 (0\%)             & 3 (21\%)       & \textbf{2 (14\%)}   & 7 (50\%)        & 5 (36\%)          & 0 (0\%)       \\
\revision{iFixitAndroid}              & 13                                          & 0 (0\%)       & 13 (100\%)          & 0 (0\%)           & 0 (0\%)          & 0 (0\%)                    & 10 (77\%)         & 3 (23\%)          & 0 (0\%)       \\
\revision{MozStumbler}                  & 10                                      & 0 (0\%)            & 10 (100\%)      & 0 (0\%)            & 0 (0\%)           & \textbf{1 (10\%)}         & 7 (70\%)          & 2 (20\%)          & 0 (0\%)       \\
\revision{error-prone}                            & 9                                         & \textbf{1 (11\%)}          & 8 (88\%)         & 0 (0\%)            & 0 (0\%)           & \textbf{1 (11\%)} & 8 (88\%)        & 0 (0\%)             & 0 (0\%)         \\

\midrule
Total                                    & 2,001     & 70 (3\%)           & 1,017 (51\%)                         & 7 (0\%)          & 907 (45\%) & \textbf{122 (6\%)} & 592 (30\%)  & 905 (45\%) & 382 (19\%)       \\ \bottomrule
\end{tabular}
}
\vspace{-4mm}
\end{table*}

In this RQ, we compare the effectiveness of each tool in resolving merge conflicts at all granularity levels: complete merge scenarios, conflicting files, conflict blocks, and conflicting lines of code reported by each merge tool for the merge scenarios in our data set.
We first focus on comparing the number of completely resolved conflicting scenarios.
Completely resolving a conflicting scenario is the best case for any tool since this relieves the developer from looking at this scenario.
While a tool may not be able to completely resolve a scenario, it may be able to reduce the number of conflicting files or conflicting regions a developer needs to deal with, or it may also reduce the size of the reported conflicts in terms of lines of code (LOC).
We report the cases in which such reduction happens.
Alternatively, a tool may worsen the situation for a developer where it complicates the conflict by reporting more conflicting files, blocks, or lines of code.
We first report detailed results of the evaluation in Sections~\ref{sec:resolved-scenarios}-\ref{sec:changed-scenarios} and then provide an interpretation of these results in Section~\ref{sec:rq1-interpretation}.

\begin{figure}
\centering
\includegraphics[width=8cm]{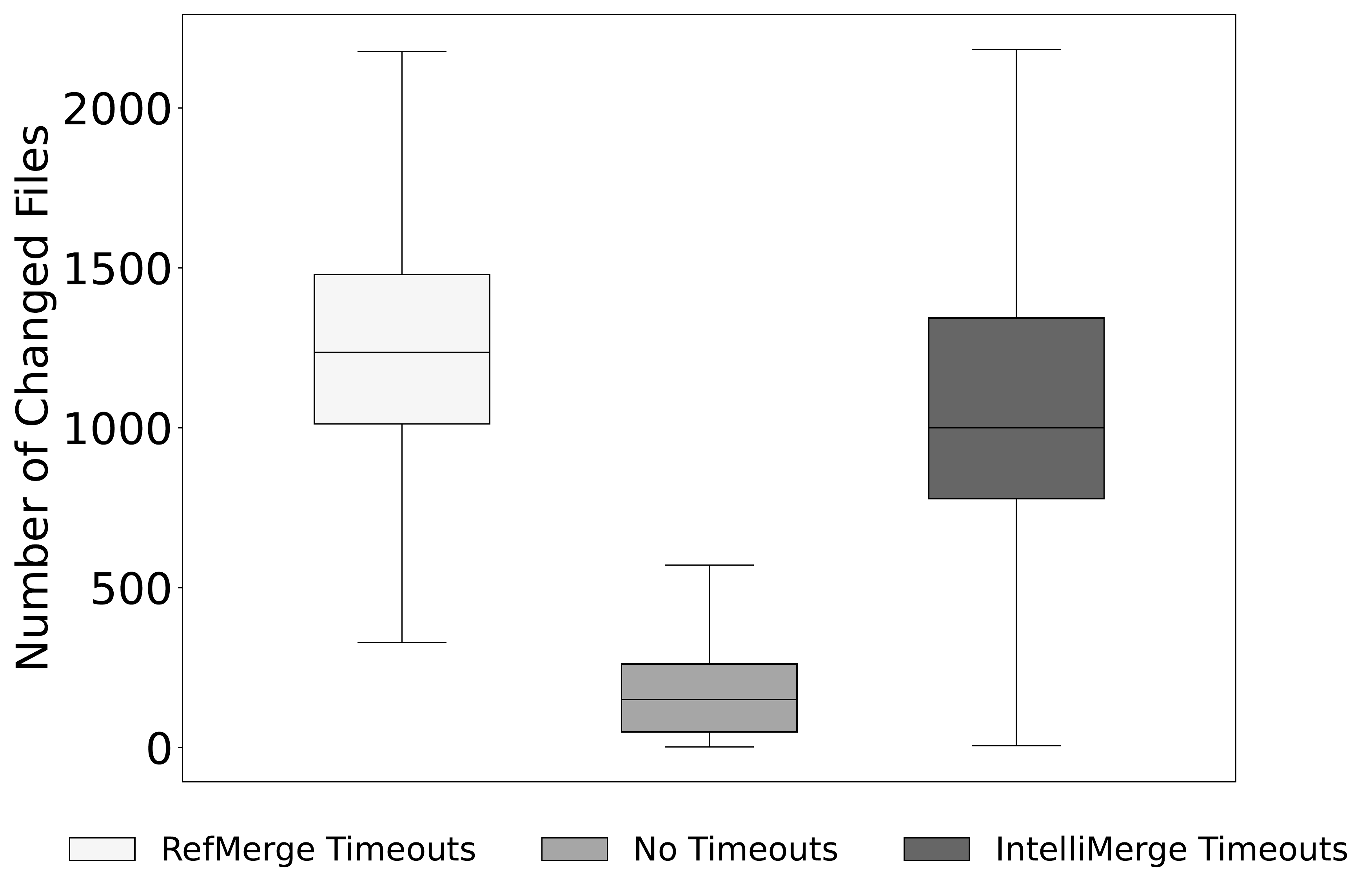}
\caption{\revision{Distribution of changed files per merge scenario where timeouts occur in each tool.} 
}
\label{fig:timeout-distribution}
\vspace{-4mm}
\end{figure}

\subsection{Completely Resolved Merge Scenarios}
\label{sec:resolved-scenarios}

Table~\ref{tab:scenario-breakdown} shows the breakdown of the merge results for each project.
The \textit{Total Scenarios} column shows the number of conflicting \git scenarios with involved refactorings evaluated for each project (same as Column c from Table~\ref{tab:conflict-scenarios}).
We then show the results for \intellimerge and \ourtool, respectively.
For each tool, we show the number of completely resolved merge scenarios (column \textit{Resolved}), the number of merge scenarios where the conflict result changed from what \git reports (column \textit{Changed}), the number of merge scenarios where the merge conflict remains the same as \git (column \textit{Unchanged}), and the number of merge scenarios where the tool times out (column \textit{Timeout}).
Note that a change in the conflict result could mean either a decrease or increase in the number or size of the reported conflicts; we discuss the details of these changed scenarios in Section~\ref{sec:changed-scenarios}.

\revision{As Table~\ref{tab:scenario-breakdown} shows, across all evaluated merge scenarios, \intellimerge was able to completely resolve \checkNum{70} merge scenarios out of the \sample total scenarios (i.e., \checkNum{3\%}) while \ourtool was able to completely resolve \checkNum{122} (\checkNum{6\%}) scenarios. 
The number of merge scenarios each tool completely resolves indicates which tool can fully resolve more scenarios. However, not all scenarios have the potential to be fully resolved. 
The strengths of refactoring-aware merge tools is their ability to deal with refactoring conflicts.
Thus, merge scenarios with \textit{only} refactoring related conflicts have more potential to be fully resolved by a refactoring-aware merge tool.
Of the \sample total conflicting scenarios, 453 (23\%) scenarios contain \textit{only} refactoring related conflicts.
All \checkNum{122} of \ourtool's resolved scenarios are a subset of these 453 scenarios, whereas only \checkNum{35} of \intellimerge's are.
This means that \ourtool resolves \checkNum{27\%} of such merge scenarios, whereas \intellimerge resolves only \checkNum{8\%}, in addition to \checkNum{35} merge scenarios with other types of conflicts.

Looking at Table~\ref{tab:scenario-breakdown} from a project perspective, \intellimerge and \ourtool are each able to completely resolve at least one scenario in \checkNum{17} \checkNum{(85\%)} projects.
Although both tools resolve merge scenarios in \checkNum{17} projects, there are \checkNum{four} projects where \intellimerge resolves more merge scenarios than \ourtool, while there are \checkNum{11} projects where \ourtool resolves more scenarios than \intellimerge. This suggests that the characteristics of the scenarios in each project play a role in the tools' capabilities in resolving them.

We note that \ourtool times out on \checkNum{382} merge scenarios across \checkNum{three} different projects and \intellimerge times out on \checkNum{870} merge scenarios across 11 projects.
To investigate the characteristics of the merge scenarios with time outs, Figure~\ref{fig:timeout-distribution} shows the distribution of changed files per merge scenario where \ourtool times out, \intellimerge times out, or neither tool times out. 
The median number of changed files in a merge scenario where \ourtool times out and \intellimerge times out are \checkNum{1,236} and \checkNum{1,000}, respectively.
In all other merge scenarios where neither tool times out, the median number of changed files is \checkNum{150} files.
Thus, it seems that merge scenarios with a large number of changed files causes both tools to time out.
This makes sense because as more files are changed, \intellimerge needs to build more, and potentially larger, graphs and then match them, which leads to it taking more time in these steps. 
As for \ourtool, more changed files adds more work for RefactoringMiner which causes \ourtool to take more time in refactoring detection. 
Overall, \ourtool and \intellimerge respectively time out on merge scenarios with \checkNum{eight} times and \checkNum{seven} times more changed files than those they do not time out on.

}

\subsection{Merge Scenarios with Differences in Conflicts}
\label{sec:changed-scenarios}

\begin{figure}
\centering
\begin{subfigure}{0.2\linewidth}
\centering
\includegraphics[width=2cm]{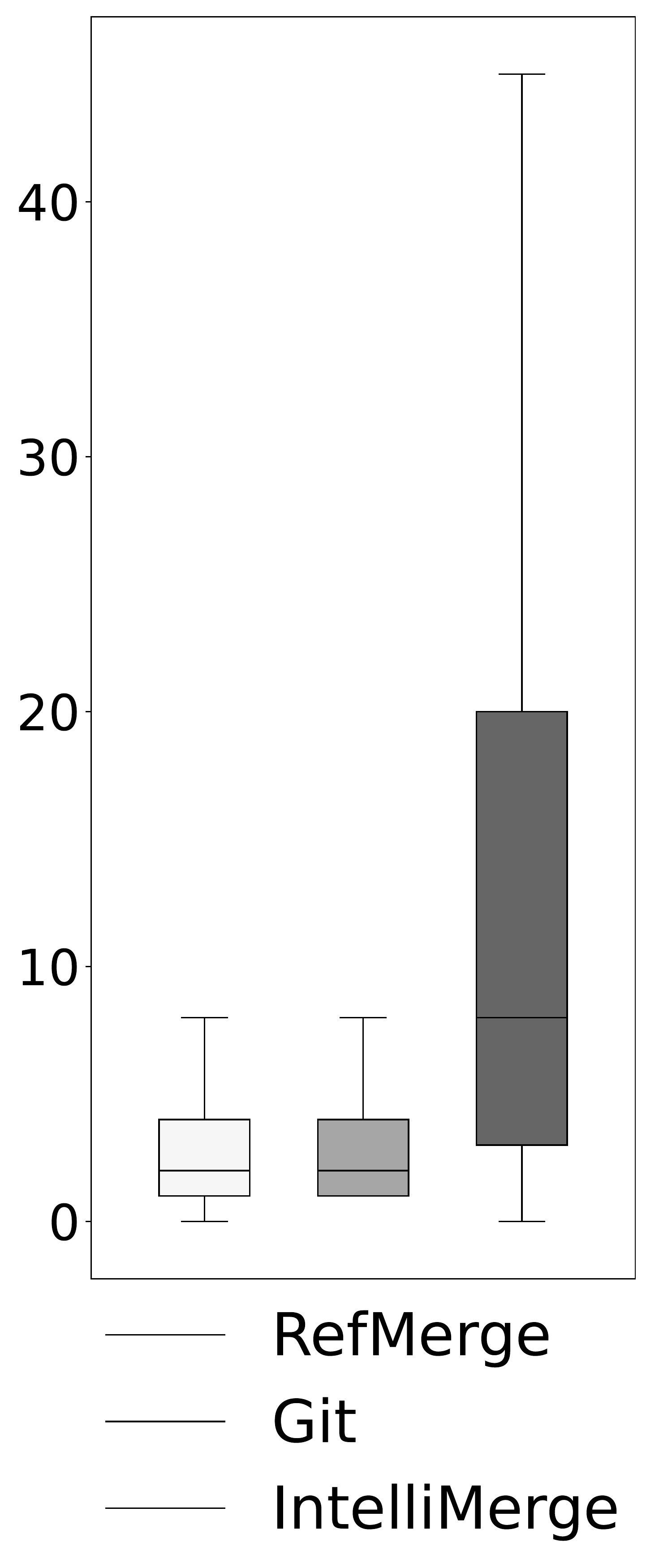}
		\caption{Overall Distribution}
		\label{fig:file-overall}
\end{subfigure}
\hspace{0.1em}
\begin{subfigure}{0.35\textwidth}
\resizebox{\textwidth}{!}{
\begin{tabular}{@{}lrrrr@{}}
\toprule
\multirow{2}{*}{\textbf{Project}} & \multicolumn{2}{c}{\textbf{Reduced Confl. Files}}                           & \multicolumn{2}{c}{\textbf{Increased Confl. Files}}                                            \\
\cmidrule(l){2-3}
\cmidrule(l){4-5}
  & \multicolumn{1}{l}{\textbf{IntelliMerge}} & \textbf{RefMerge}  & \multicolumn{1}{l}{\textbf{IntelliMerge}} & \multicolumn{1}{l}{\textbf{RefMerge}} \\
  \midrule
\revision{cassandra}                     & 1 (33\%)                                   &  \textbf{34 (27\%)} &  105 (567\%)                                & \textbf{33 (50\%)}                   \\
\revision{elasticsearch}                 & 4 (20\%)                                  & \textbf{14 (21\%)} & 99 (260\%)                                & \textbf{9 (67\%)}                    \\
\revision{gradle}                        & \textbf{7 (33\%)}                                  & \textbf{7 (25\%)} & 86 (494\%)                                & \textbf{4 (100\%)}                    \\
\revision{antlr4}                        & 2 (56\%)                                  & \textbf{11 (17\%)}  & 92 (445\%)                                & \textbf{4 (16\%)}                    \\
\revision{platform\_fwk\_supp} & \textbf{6 (68\%)}                         & 5 (50\%)            & 45 (350\%)                                & \textbf{4 (14\%)}                     \\
\revision{deeplearning4j}                & \textbf{6 (67\%)}                         & 3 (50\%)            & 61 (150\%)                                & \textbf{9 (50\%)}                     \\
\revision{realm-java}                    & \textbf{11 (33\%)}                        & 7 (33\%)           & 45 (200\%)                                & \textbf{6 (58\%)}                     \\
\revision{jackson-core}                  & 0 (0\%)                                   & \textbf{6 (29\%)}  & 79 (850\%)                                & \textbf{0 (0\%)}                    \\
\revision{android}                       & \textbf{6 (38\%)}                         & 4 (13\%)           & 55 (150\%)                                & \textbf{5 (23\%)}                     \\
\revision{cometd}                        & 2 (48\%)                                  & \textbf{3 (25\%)}  & 48 (658\%)                                & \textbf{4 (15\%)}                      \\
\revision{storm}                         & \textbf{3 (33\%)}                         & 2 (42\%)           & 21 (340\%)                                & \textbf{0 (0\%)}                    \\
\revision{ProjectE}                      & \textbf{4 (12\%)}                         & 3 (8\%)           & 24 (79\%)                                 & \textbf{4 (17\%)}                      \\
\revision{javaparser}                    & \textbf{5 (43\%)}                         & 4 (22\%)           & 8 (150\%)                                 & \textbf{1 (100\%)}                     \\
\revision{druid}                         & \textbf{7 (43\%)}                         & 1 (29\%)           & \textbf{0 (0\%)}                          & \textbf{0 (0\%)}                              \\
\revision{androidannotations}            & 1 (67\%)                                  & \textbf{2 (30\%)}  & 13 (100\%)                                & \textbf{0 (0\%)}                     \\
\revision{junit4}                        & \textbf{5 (33\%)}                         & 2 (42\%)            & 7 (150\%)                                 & \textbf{4 (42\%)}                     \\
\revision{MinecraftForge}                & 1 (10\%)                                  & \textbf{1 (17\%)}  & 9 (100\%)                                 & \textbf{1 (11\%)}                    \\
\revision{iFixitAndroid}                 & \textbf{5 (81\%)}                         & 6 (20\%)           & 5 (78\%)                                  & \textbf{1 (33\%)}                   \\
\revision{MozStumbler}                   & 1 (25\%)                                  & \textbf{4 (33\%)}  & 5 (50\%)                                  & \textbf{0 (0\%)}                     \\
\revision{error-prone}                   & \textbf{4 (46\%)}                         & 3 (25\%)           & 3 (750\%)                                 & \textbf{0 (0\%)}                      \\
\midrule
Total                         & 81 (38\%)                                 & \textbf{122 (25\%)} & 810 (350\%)                     &  \textbf{89 (50\%)}                   \\ \bottomrule
\end{tabular}}
\caption{Breakdown by merge scenario}
\label{fig:file-change}
\end{subfigure}
\caption{Conflicting \textit{files} in merge scenarios. Boxplot shows number of conflicting files per merge scenario while the table shows number of merge scenarios where a tool reduced/increased the number of conflicting files, compared to Git. In parenthesis, we show the median percentage reduction/increase per merge scenario. 
}
\label{fig:files}
\vspace{-5mm}
\end{figure}

\begin{figure}
\centering
\begin{subfigure}{0.2\linewidth}
\centering
\includegraphics[width=2cm]{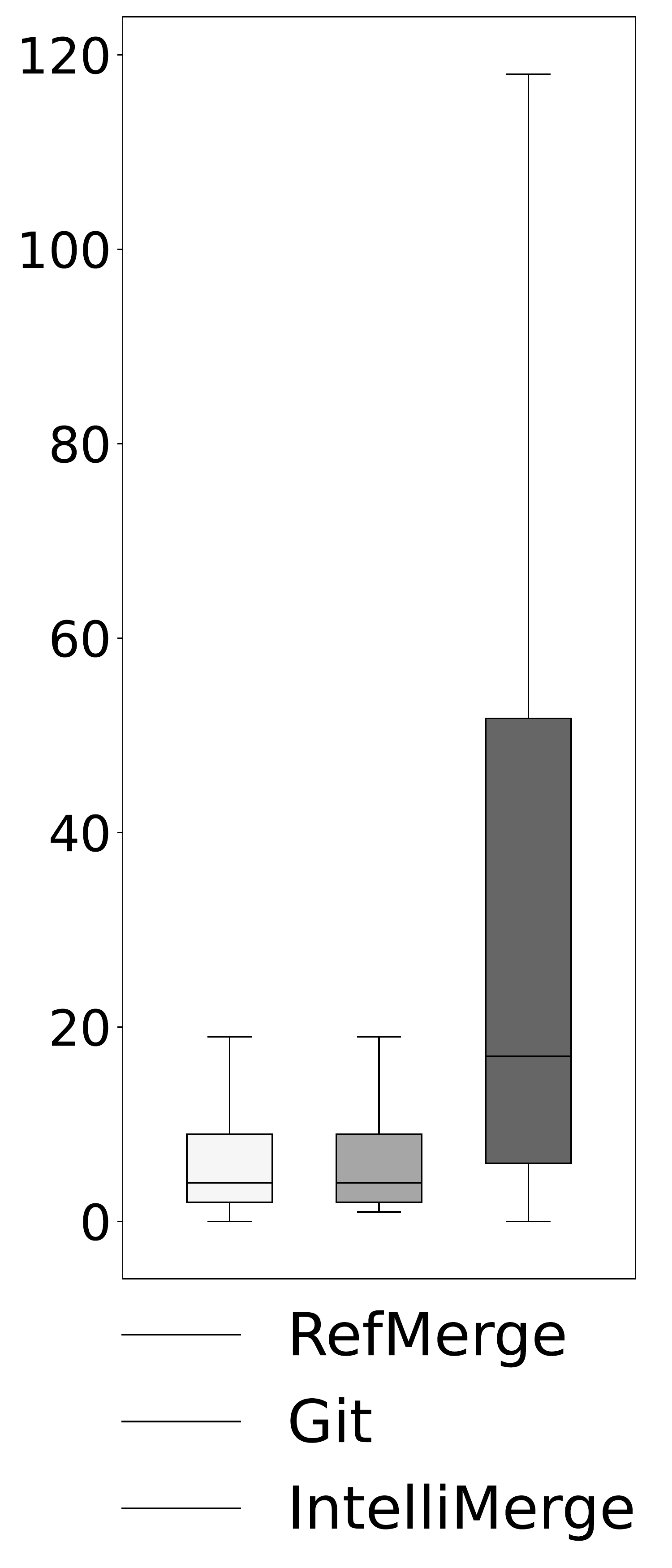}
		\caption{Overall Distribution}
		\label{fig:block-overall}
\end{subfigure}
\hspace{0.1em}
\begin{subfigure}{0.35\textwidth}
\resizebox{\textwidth}{!}{
\begin{tabular}{@{}lrrrr@{}}
\toprule
\multirow{2}{*}{\textbf{Project}} & \multicolumn{2}{c}{\textbf{Reduced Confl. Blocks}}                           & \multicolumn{2}{c}{\textbf{Increased Confl. Blocks}}                                            \\
\cmidrule(l){2-3}
\cmidrule(l){4-5}
  & \multicolumn{1}{l}{\textbf{IntelliMerge}} & \textbf{RefMerge}  & \multicolumn{1}{l}{\textbf{IntelliMerge}} & \multicolumn{1}{l}{\textbf{RefMerge}} \\
  \midrule
\revision{cassandra}              & 4 (29\%)                                   & \textbf{68 (25\%)}    & 101 (767\%)                                  & \textbf{58 (27\%)}            \\
\revision{elasticsearch}          & 5 (20\%)                                  & \textbf{19 (14\%)}  & 94 (294\%)                                  & \textbf{13 (29\%)}            \\
\revision{gradle}                 & \textbf{10 (29\%)}                                 & \textbf{10 (25\%)}  & 91 (400\%)                                  & \textbf{14 (50\%)}            \\
\revision{antlr4}                 & 3 (20\%)                                  & \textbf{16 (8\%)}    & 92 (479\%)                                  & \textbf{7 (9\%)}            \\
\revision{platform\_fwk\_supp} & 9 (50\%)                                  & \textbf{13 (20\%)}   & 47 (508\%)                                  & \textbf{9 (13\%)}            \\
\revision{deeplearning4j}         & \textbf{18 (44\%)}                                 & 13 (11\%)    & 65 (167\%)                                  & \textbf{10 (33\%)}            \\
\revision{realm-java}             & \textbf{22 (50\%)}                                 & 20 (28\%)  & 47 (200\%)                                  & \textbf{5 (23\%)}             \\
\revision{jackson-core}           & 0 (0\%)                                   & \textbf{13 (17\%)}  & 80 (1580\%)                                 & \textbf{3 (19\%)}            \\
\revision{android}                & \textbf{20 (39\%)}                                 & 11 (20\%)   & 52 (119\%)                                  & \textbf{8 (32\%)}             \\
\revision{cometd}                 & 5 (26\%)                                  & \textbf{7 (33\%)}   & 52 (790\%)                                  & \textbf{6 (21\%)}             \\
\revision{storm}                  & 3 (25\%)                                  & \textbf{4 (29\%)}   & 25 (233\%)                                  & \textbf{0 (0\%)  }            \\
\revision{ProjectE}               & 4( 15\%)                         & \textbf{5 (8\%)}   & 22 (94\%)                                   & \textbf{5 (11\%) }            \\
\revision{javaparser}             & 3 (89\%)                                  & \textbf{5 (12\%)}   & 11 (138\%)                                  & \textbf{1 (100\%) }            \\
\revision{druid}                  & \textbf{12 (60\%)}                                 & 3 (8\%)   & \textbf{0 (0\%)}                                     & 1 (100\%)              \\
\revision{androidannotations}     & 1 (89\%)                                  & \textbf{5 (33\%)}   & 13 (114\%)                                  & \textbf{1 (25\%)}             \\
\revision{junit4}                 & \textbf{5 (60\%)}                                  & 4 (32\%)    & 7 (233\%)                                   &\textbf{3 (20\%)}             \\
\revision{MinecraftForge}         & 1 (5\%)                                   & \textbf{3 (10\%)}   & 9 (100\%)                                   & \textbf{3 (20\%)}             \\
\revision{iFixitAndroid}          & 6 (57\%)                                  & \textbf{8 (18\%)}   & 7 (100\%)                                   & \textbf{1 (50\%)}            \\
\revision{MozStumbler}            & 3 (50\%)                                  & \textbf{6 (20\%)}   & 5 (86\%)                                    & \textbf{0 (0\%)}             \\
\revision{error-prone}            & 2 (91\%)                                  & \textbf{5 (33\%)}   & 5 (417\%)                                   & \textbf{0 (0\%)}              \\
\midrule
Total                  & 136 (48\%)                                & \textbf{239 (21\%)} & 825 (378\%)                         & \textbf{149 (27\%)}           \\ \bottomrule
\end{tabular}}
\caption{Breakdown by merge scenario}
\label{fig:block-change}
\end{subfigure}
\caption{Conflicting \textit{blocks} in merge scenarios. Boxplot shows number of conflicting blocks per merge scenario while the table shows number of merge scenarios where a tool reduced/increased the number of conflicting blocks, compared to Git. In parenthesis, we show the median percentage reduction/increase per merge scenario. 
}
\label{fig:blocks}
\vspace{-5mm}
\end{figure}

We now look at the \textit{remaining} scenarios that the tools are not able to completely resolve, but for which the result of the conflict changed.
We use Figures~\ref{fig:files}-\ref{fig:loc} to discuss these scenarios per project at the file, block, and lines of code levels respectively\footnote{We use platform\_fwk\_supp as a shortened version of platform\_frameworks\_support for better table sizing for readability}.
There are two parts to each figure.
On the left-hand side, we provide a box plot of the overall distribution of reported conflicts at the discussed granularity level for all three tools across all evaluated scenarios.
On the right-hand side, we show a table that provides the details of the conflicting scenarios from the \textit{Changed} column of Table~\ref{tab:scenario-breakdown}.
For each granularity level (conflicting files, conflict blocks, and conflict size in terms of LOC), we show the number of scenarios for which a tool increased or decreased the resulting number of conflicts.
For example, for the last project \texttt{error-prone} in Figure~\ref{fig:file-change}, we can see that there are \checkNum{four} scenarios that \intellimerge reduced the number of conflicting files for while it increased the number of conflicting files for \checkNum{three} scenarios.
The percentage shown in parentheses is the median reduction/increase per merge scenario in that project (or over all scenarios in the last row of the table).
For example, if \git reports 4 conflicting files while a tool reports 2 conflicting files, then this is a $(4-2)/4 = 50\%$ reduction.
In the example of \texttt{error-prone}, the median reduction of the number of conflicting files for the corresponding \checkNum{four} scenarios is \checkNum{46\%}.
The same interpretation of the numbers can be used for all granularity levels, which we discuss in detail below.
Ideally, even if a tool cannot completely resolve a scenario, it would be able to partially resolve some of the reported conflicts.
For each project, we show in bold which tool achieves the \textit{most} reduction and the \textit{least} increase.

\paragraph*{\textbf{\textit{Conflicting files}}}
We first look at the conflicting file level in Figure~\ref{fig:files}.
Figure~\ref{fig:file-overall} shows the distribution of the number of reported conflicting files per merge scenario.
The figure shows that \git and \ourtool have a median number of two conflicting files while \intellimerge has a median of eight.
However, such a plot does not give us any indication about the developer experience on a scenario level, when it compares to what they currently experience with \git.
To understand the tool's behavior on a scenario level, we look at the table in Figure~\ref{fig:file-change}, which shows the number of scenarios for which each tool results in an increase or decrease in the number of conflicting files.
\revision{
Overall, the table shows that \intellimerge reduces the number of reported conflicting files in \checkNum{81} scenarios (\checkNum{4\%} of all evaluated scenarios) by a median \checkNum{38\%} reduction.
On the other hand, \intellimerge increases the number of reported conflicting files in \checkNum{810} scenarios (\checkNum{40\%}) by a median \checkNum{350\%} increase.
In other words, on average, \intellimerge increases the number of conflicting files by three-fold in these scenarios.
\ourtool reduces the number of reported conflicting files for \checkNum{122} scenarios (\checkNum{6\%}) by a median \checkNum{25\%} reduction while it increases the number of reported conflicting files for \checkNum{89} (\checkNum{4\%}) by a median \checkNum{50\%} increase.

}

\paragraph*{\textbf{\textit{Conflict Blocks}}}
We now look at the conflict block level in Figure~\ref{fig:block-change}.
The number of conflict blocks indicates the number of individual conflicting regions a developer needs to deal with.
\revision{
Figure~\ref{fig:block-overall} shows that \git and \ourtool have almost the same overall distribution of number of conflicting blocks per merge scenario (with a median of \checkNum{4}).
However, \intellimerge has a much higher median number of conflicting blocks at \checkNum{17}.
From Figure~\ref{fig:block-change}, we find that \intellimerge reduces the number of reported conflict blocks for \checkNum{136} scenarios (\checkNum{7\%}) by a median \checkNum{48\%} reduction, while it increases the number of reported conflict blocks for \checkNum{825} scenarios (\checkNum{41\%}) by a median of \checkNum{378\%}.
On the other hand, \ourtool reduces the number of reported conflicts in \checkNum{239} scenarios (\checkNum{12\%}) by a median \checkNum{21\%} reduction and increases the number of reported conflicts for \checkNum{197} scenarios (\checkNum{7\%}) by a median of \checkNum{27\%} increase.
Additionally, \intellimerge has a high variance between projects. For example, consider project \textit{druid}. \intellimerge reduces conflict blocks in 12 merge scenarios by more than half (60\%) without increasing them in any merge scenarios. 
Inversely, it increases conflict blocks in 80 merge scenarios for project \textit{jackson-core} by almost 16-fold without decreasing conflict blocks in any merge scenarios. 
To investigate the cause of the variance, we look at Figure~\ref{fig:refactoring-distribution}.
As shown, \textit{druid} has the lowest median number of refactorings in each scenario, while the four projects that \intellimerge struggles most with (\textit{cassandra}, \textit{antlr4}, \textit{jackson-core}, and \textit{cometd}) have the highest median number of refactorings. 
This suggests that merge scenarios with a large number of refactorings complicates the merge resolution for \intellimerge.
\ourtool does not exhibit the same variance as \intellimerge, typically reducing conflicts more often than increasing conflicts or reducing them as often as it increases them. 
To explore how the number of refactorings performed in a project affects \ourtool, we look at \textit{MozStumbler}, \textit{error-prone}, and \textit{antlr4} because \ourtool does well on them.
We additionally look at \textit{gradle} because \ourtool does the worst on it. 
\textit{MozStumbler} has the third lowest median number of refactorings per scenario while \textit{antlr4} has the second highest median number of refactorings per scenario.
Projects \textit{error-prone} and \textit{gradle} respectively have the 9th and 8th highest median number of refactorings per scenario.
This suggests that \ourtool is not affected by the number of refactorings in a project.
}

\paragraph*{\textbf{\textit{Conflicting Lines of Code}}}

Finally, we look at the conflicting lines of code (LOC) in Figure~\ref{fig:loc}, which measures the total number of lines in all conflict blocks/regions of a merge scenario.
From Figure~\ref{fig:loc-overall}, we observe similar behavior of the tools as what we observed for the conflicting files in Figure~\ref{fig:file-overall}.
\revision{
More closely from the table in Figure~\ref{fig:loc-change}, we find that \intellimerge reduces the number of conflicting LOC in \checkNum{408} scenarios \checkNum{(20\%)} by a median \checkNum{51\%} reduction, while it increases the conflicting LOC for \checkNum{597 (30\%)} scenarios by a median \checkNum{169\%} increase.
\ourtool reduces the conflicting LOC in \checkNum{372} scenarios (\checkNum{19\%}) by a median \checkNum{21\%} reduction and increases the conflicting LOC in \checkNum{214} scenarios (\checkNum{11\%}) by a median \checkNum{14\%} increase.
Note that the discrepancy between \intellimerge's increased rate for conflicting regions and conflicting LOC suggests that while \intellimerge results in a lot more conflicting regions than \git, the size of these conflicting regions is small.
To confirm this, we show the distribution of the reported conflicting LOC per block (as opposed to a whole scenario) in Figure~\ref{fig:locPerBlock}.
The plot confirms that the conflict regions that \intellimerge reports are indeed quite small, even if they are much more frequent than the other tools.
}

\subsection{Summary and Interpretation of RQ1 Results}
\label{sec:rq1-interpretation}

\begin{figure}
\centering
\begin{subfigure}{0.2\linewidth}
\centering
\includegraphics[width=2cm]{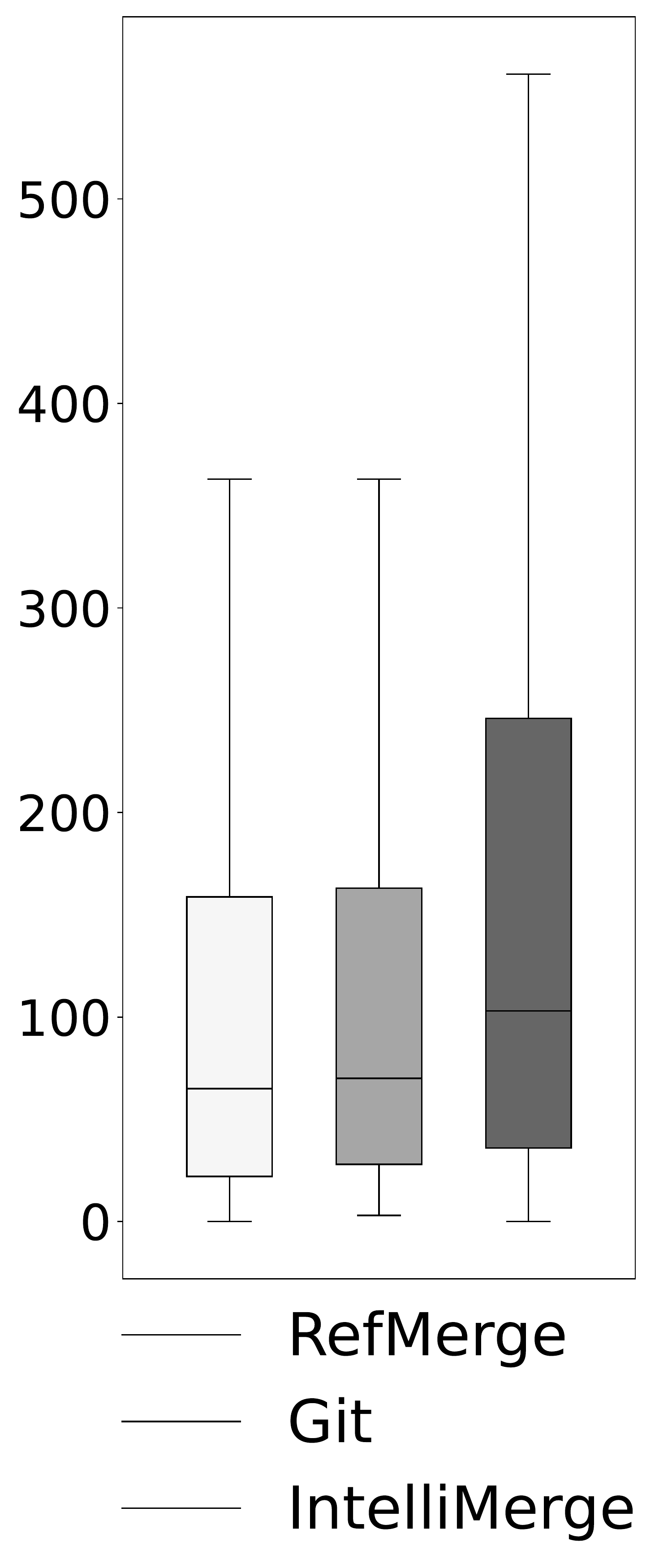}
		\caption{Overall Distribution}
		\label{fig:loc-overall}
\end{subfigure}
\hspace{0.1em}
\begin{subfigure}{0.35\textwidth}
\resizebox{\textwidth}{!}{
\begin{tabular}{@{}lrrrr@{}}
\toprule
\multirow{2}{*}{\textbf{Project}} & \multicolumn{2}{c}{\textbf{Reduced Confl. LOC}}                           & \multicolumn{2}{c}{\textbf{Increased Confl. LOC}}                                            \\
\cmidrule(l){2-3}
\cmidrule(l){4-5}
  & \multicolumn{1}{l}{\textbf{IntelliMerge}} & \textbf{RefMerge}  & \multicolumn{1}{l}{\textbf{IntelliMerge}} & \multicolumn{1}{l}{\textbf{RefMerge}} \\
  \midrule
\revision{cassandra}              & 23 (36\%)                          & \textbf{124 (26\%)}                           & 85 (291\%)                          & \textbf{80 (14\%}                             \\
\revision{elasticsearch}          & \textbf{39 (46\%)}                         & 28 (14\%)                         & 64 (121\%)                          & \textbf{16 (28\%)}                           \\
\revision{gradle}                 & \textbf{37 (45\%)}                         & 20 (26\%)                         & 69 (149\%)                          & \textbf{11 (61\%)}                           \\
\revision{antlr4}                 & 19 (30\%)                         & \textbf{21 (6\%)}                          & 75 (102\%)                          & \textbf{18 (8\%)}                           \\
\revision{platform\_fwk\_supp} & \textbf{23 (59\%)}                         & 24 (20\%)                         & 33 (129\%)                          & \textbf{16 (11\%)}                           \\
\revision{deeplearning4j}         & \textbf{45 (45\%)}                          & 16 (14\%)                          & 41 (148\%)                    & \textbf{15 (11\%)}                                  \\
\revision{realm-java}             & \textbf{53 (74\%)}                         & 25 (20\%)                         & 29 (68\%)                          & \textbf{7 (25\%)}                           \\
\revision{jackson-core}           & 8 (48\%)                          & \textbf{20 (14\%)}                          & 72 (420\%)                          & \textbf{6 (10\%)}                           \\
\revision{android}                & \textbf{54 (58\%)}                         & 18 (15\%)                          & 23 (119\%)                          & \textbf{9 (14\%)}                            \\
\revision{cometd}                 & \textbf{15 (51\%)}                         & 9 (36\%)                          & 43 (230\%)                          & \textbf{11 (9\%)}                            \\
\revision{storm}                  & \textbf{14 (39\%)}                         & 12 (18\%)                          & 14 (113\%)                          & \textbf{1 (3\%)}                             \\
\revision{ProjectE}               & \textbf{18 (45\%)}                         & 8 (3\%)                         & 10 (75\%)                           & \textbf{5 (9\%)}                             \\
\revision{javaparser}             & \textbf{11 (59\%)}                         & 8 (24\%)                          & \textbf{8 (225\%)}                           & \textbf{0 (0\%)}                            \\
\revision{druid}                  & \textbf{14 (85\%)}                         & 6 (19\%)                          & \textbf{0 (0\%)}                             & 2 (32\%)                            \\
\revision{androidannotations}     & 4 (87\%)                          & \textbf{8 (19\%)}                          & 10 (42\%)                           & \textbf{1 (7\%)}                            \\
\revision{junit4}                 &\textbf{ 10 (65\%)}                         & 5 (20\%)                           & \textbf{3 (18\%)}                            & 4 (8\%)                            \\
\revision{MinecraftForge}         & \textbf{4 (27\%)}                          & 3 (11\%)                 & \textbf{4 (150\%)}                           & \textbf{4 (16\%)}                            \\
\revision{iFixitAndroid}          & \textbf{7 (34\%)}                          & 5 (13\%)                          & \textbf{5 (55\%)}                            & \textbf{5 (21\%)}                            \\
\revision{MozStumbler}            & \textbf{7 (60\%)}                          & 6 (31\%)                          & 3 (38\%)                            & \textbf{1 (12\%)}                            \\
\revision{error-prone}            & 3 (84\%)                          & \textbf{6 (47\%)}                          & 5 (187\%)                         & \textbf{2 (25\%)}                            \\
\midrule
Total                  & \textbf{408 (51\%)}                        & 372 (21\%)                        & 597 (169\%)                         & \textbf{214 (14\%)}                          \\ \bottomrule
\end{tabular}}
\caption{Breakdown by merge scenario}
\label{fig:loc-change}
\end{subfigure}
\caption{Conflicting \textit{LOC} in merge scenarios .Boxplot shows number of conflicting LOC per merge scenario while the table shows number of merge scenarios where a tool reduced/increased the number of conflicting LOC, compared to Git. In parenthesis, we show the median percentage reduction/increase per merge scenario.
}
\label{fig:loc}
\vspace{-5mm}
\end{figure}

\revision{
The above results indicate that \ourtool completely resolves almost twice as many merge scenarios as \intellimerge (\checkNum{122 versus 70}).
 Of the 453 merge scenarios with \textit{only} refactoring-related conflicts, \intellimerge and \ourtool respectively resolve \checkNum{8\%} and \checkNum{27\%}.

While \intellimerge is able to reduce conflicting LOC for a higher portion of scenarios than \ourtool (\checkNum{51\% versus \checkNum{21\%}}), this comes at a cost of a high increase in the reported conflicts across all granularity levels for a large portion of the merge scenarios. 
Furthermore, there is a large variance in how \intellimerge does across the projects. 
We find that \intellimerge does well on projects whose merge scenarios have a low number of refactorings and struggles with projects that have several refactorings per merge scenario.
Additionally, \intellimerge times out on a higher number of merge scenarios than \ourtool. 
In our investigation into why each tool times out, we found that merge scenarios with more changed files typically cause both tools to time out. 
Thus, it seems \intellimerge works extremely well for merge scenarios with few changes and a small number of refactorings, but actually makes it much worse for other scenarios. 
This makes sense because \intellimerge relies on a similarity score to detect refactorings and a large number of changes make it harder to detect refactorings.
Overall, taking the total number of scenarios it can completely resolve (\checkNum{70} from Table~\ref{tab:scenario-breakdown}) and the ones in which it can reduce the total number of conflicting LOC for (\checkNum{408} from Figure~\ref{fig:loc-change}), \intellimerge can help the developer deal with less conflicts in \checkNum{478} scenarios (\checkNum{24\%} of overall scenarios).
However, taking both timeouts (\checkNum{907} scenarios from Table~\ref{tab:scenario-breakdown}) and worsened results in terms of overall conflicting LOC (\checkNum{597} scenarios from Figure~\ref{fig:loc-change}), \intellimerge will not help the developer in the remaining \checkNum{1,504} (\checkNum{75\%}) scenarios, and will in fact make it worse for them in almost a third of those.

\begin{figure}[t!]
    \centering
    \includegraphics[width=0.33\textwidth]{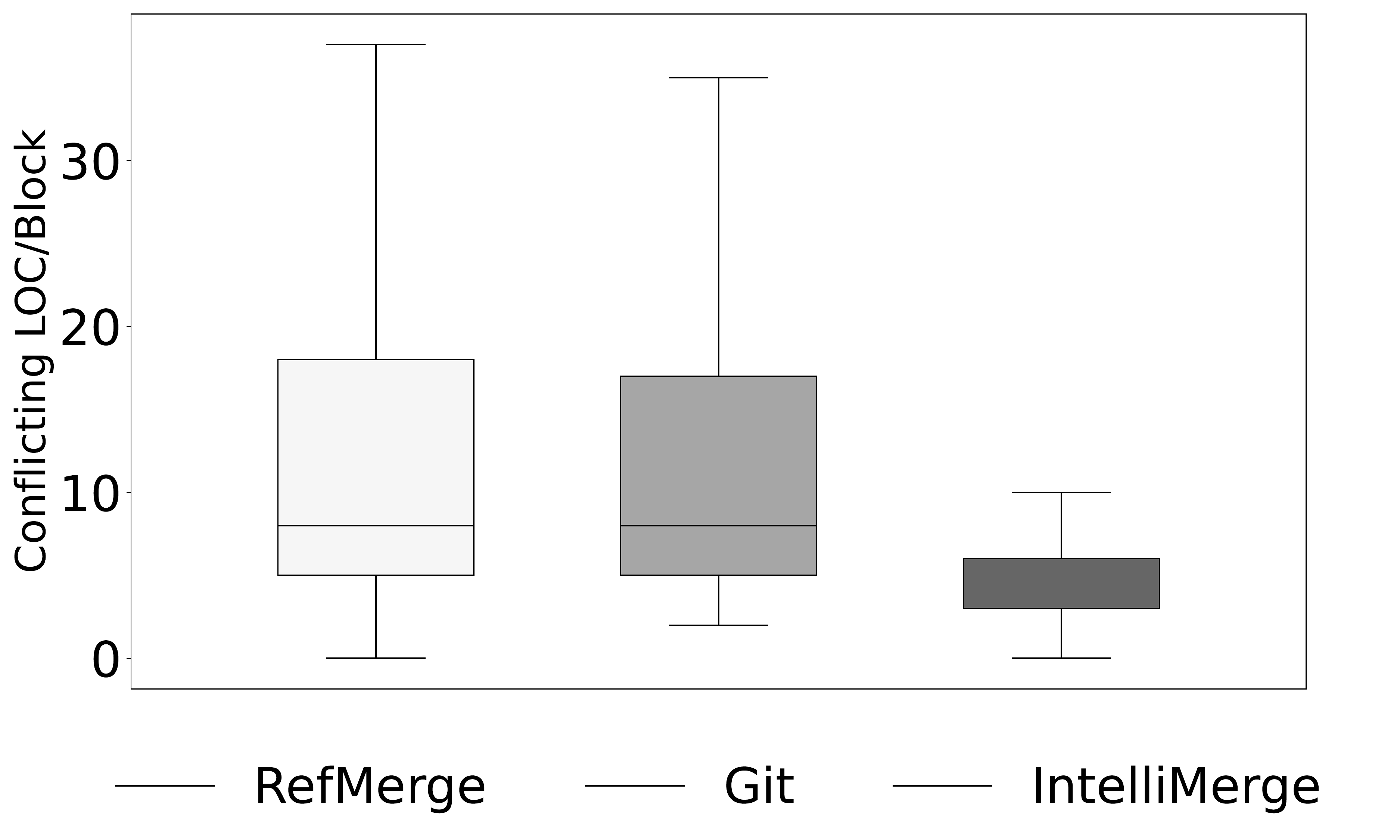}
    \caption{Conflicting lines of code per conflict block.}
    \label{fig:locPerBlock}
\end{figure}

On the other hand, \ourtool completely resolves or reduces the number of conflicting LOC for \checkNum{497} scenarios (\checkNum{25\%} of overall scenarios).
Thus, \ourtool helps the developer in around the same number of merge scenarios as \intellimerge but it times out or worsens the situation at a much lower rate, only \checkNum{596 (30\%)}.
Additionally, the median percentage increase for \ourtool in terms of conflicting LOC is much lower at \checkNum{14\%} as opposed to \checkNum{169\%} for \intellimerge.
Thus, \ourtool makes the situation worse for the developer both in a smaller proportion of merge scenarios \textit{and} by a lower percentage increase.
Note that the number of unchanged merge scenarios for \ourtool is also much higher than \intellimerge, because by construction, \ourtool resorts to a regular \git merge when there are no supported refactorings for it to work with.

Overall, our quantitative results show that while overall \ourtool seems to be doing better quantitatively, it is obvious that the characteristics or difficulty of a merge scenario impact the results in some way.
One tool may be able to handle certain types of merge scenarios better than the other, and we do not have information about the correctness of the resolutions.
This is why we perform a qualitative analysis of these discrepancies in RQ2 to understand the strengths and weaknesses of each tool, as well as the characteristics of merge scenarios that cause them to fail.

\begin{findingenv}{RQ1 Summary}{finding:rq1-LOC}
\ourtool completely resolves double the total number of merge scenarios as \intellimerge (122 (6\%) vs 70 (3\%)).
Overall, \intellimerge can help the developer completely resolve conflicts or deal with less conflicts, and thus improve the situation, in \checkNum{478} scenarios (\checkNum{24\%}), but at the cost of increasing conflicting LOC, thus worsening the conflicts, in \checkNum{597} (30\%) scenarios. In contrast, \ourtool improves the situation in \checkNum{497} (25\%) scenarios, and worsens it in only \checkNum{214} (11\%) scenarios by a lower LOC increase rate.



\end{findingenv}

}

\section{RQ2: Discrepancies between the tools}

RQ1 quantitative results are valuable for determining if a merge tool reports less conflicts. However, these numbers do not provide us information about the \textit{quality} of the resolutions the tools provide.
For example, a merge tool could report no conflicts in a merge scenario where conflicts should be reported. 
Similarly, the reported conflicts may not be real conflicts.
Thus, we perform a qualitative study for RQ2 to dig deeper into the reported results.

\subsection{Research Method}

\paragraph*{\textit{Sampling Criteria}} We manually analyze a sample of 50 merge scenarios to shed light on the strengths and weaknesses of each tool. 
We randomly sample the 50 merge scenarios across the following criteria: (1) \intellimerge and \ourtool produce similar results by completely resolving the merge scenario, or equally increasing/reducing the number of \git conflicts. 
(2) \intellimerge outperforms \ourtool by completely resolving the scenario or reporting a lower number of conflicts at any granularity level
and (3) \ourtool outperforms \intellimerge.
We also try to evenly sample across projects.

\paragraph*{\textit{Analysis Method}} 
Our manual analysis goal is to analyze the conflicts reported by all three tools across the sampled scenarios.
To investigate if a merge conflict is a true/false positive, we look at the code region in the base commit, left commit, and right commit. 
We determine whether integrating the changes from both parents should result in a merge conflict, based on the semantics of the changes. 
If a merge conflict is expected \revision{because it requires developer intervention}, we label this conflict region as a \textit{true positive}. 
If it should not result in a merge conflict \revision{(i.e., a tool should be able to automatically resolve it)}, we label it as a \textit{false positive}. If the other merge tools do not report the same conflict, we investigate the result of their merge and decide if it is a \textit{true negative} (i.e., conflict should not be reported) or \textit{false negative} (i.e., the other tool(s) missed the conflict).
We also investigate and categorize the reasons behind false positives and false negatives for each tool.
This process takes an average of \checkNum{63} minutes per merge scenario.

\subsection{Results}

\revision{Table~\ref{tab:additional-conflicts} shows the total number of conflict blocks that we manually analyze across the 50 sampled scenarios, as well as the number of false positives and false negatives that we find for each tool.
As shown, \git reports \checkNum{243} false positives and \checkNum{5} false negatives. 
\intellimerge reports \checkNum{707} false positives and \checkNum{71} false negatives.
Meanwhile, \ourtool reports \checkNum{188} false positives and no false negatives.
When compared to \git, \ourtool reduces the number of false positives by \checkNum{23\%} and completely eliminates false negatives, while \intellimerge increases the number of false positives and false negatives by \checkNum{192\%} and \checkNum{1,320\%}. 
We also show the number of true positives reported by each tool. 
While \git and \ourtool report a total of \checkNum{190} and \checkNum{191} true positives respectively, \intellimerge reports only \checkNum{159} true positives.
}


\begin{table}[t]
\centering
\caption{Comparing the false positives and false negatives reported by each tool, across the 50 sampled scenarios.}
\label{tab:additional-conflicts}
\hspace{0.1em}
\resizebox{0.4\textwidth}{!}{
\begin{tabular}{@{}lrrr@{}}
\toprule
 & \textbf{RefMerge} & \textbf{Git}  & \textbf{IntelliMerge} \\
\midrule
\revision{\# Conflict Blocks Investigated} &379& 433 & 866 \\
\midrule
\revision{True Positives} & 191 & 190 & 159\\
\revision{False Positives} & 188 & 243 & 707  \\
\revision{False Negatives}  &  0 & 5 & 71 \\
\bottomrule
\end{tabular}}
\end{table}

\begin{table}[t]
\centering
\caption{The reason for each false positive and false negative reported by Git, as well as the frequency for each reason.}
\label{tab:categories-git}
\hspace{0.1em}
\resizebox{0.35\textwidth}{!}{
\begin{tabular}{@{}llr@{}}
\toprule
\textbf{Git Reasons}     & \textbf{Type}     & \textbf{Frequency}       \\
\midrule 
\revision{No Refactoring Handling}    &   False Positive  &   140             \\
\revision{Ordering Conflict}       &   False Positive  &   61              \\
\revision{Formatting Conflict}     &   False Positive  &   41              \\
\revision{No Refactoring Handling}    &   False Negative  &   5              \\
\midrule
\revision{Total}& &                                        248              \\
\bottomrule
\end{tabular}}

\vspace{-4mm}
\end{table}

\begin{figure}[t]
	\centering
	\begin{subfigure}{.225\textwidth}
		\centering
		\includegraphics[width=\textwidth]{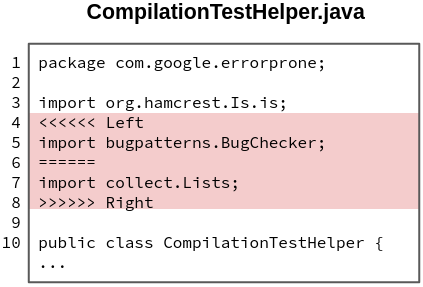}
		\caption{Git and RefMerge Result}
		\label{subfig:git-oc1}
	\end{subfigure}
		\begin{subfigure}{.225\textwidth}
		\centering
		\includegraphics[width=\textwidth]{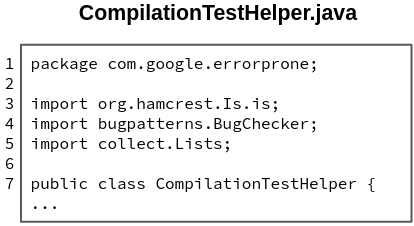}
		\caption{IntelliMerge Result}
		\label{subfig:ideal-oc1}
	\end{subfigure}
	
    \caption{\revision{An ordering conflict reported by \git and \ourtool along with the correct merge resolution by \intellimerge (error-prone [\texttt{\href{https://github.com/google/error-prone/commit/07559b47674594fdf40f2855f83b492f67f9093c}{07559b47}}]).}}
    \label{fig:ordering-conflict}
\vspace{-4mm}
\end{figure}

\begin{figure}[t]
	\centering
	\begin{subfigure}{.225\textwidth}
		\centering
		\includegraphics[width=\textwidth]{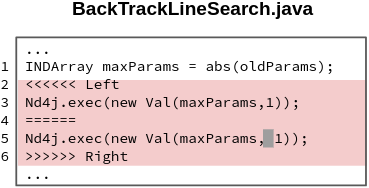}
		\caption{Git and RefMerge Result}
		\label{subfig:git-fc1}
	\end{subfigure}
		\begin{subfigure}{.225\textwidth}
		\centering
		\includegraphics[width=\textwidth]{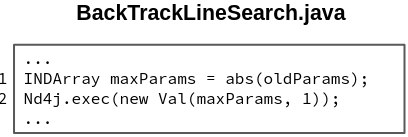}
		\caption{IntelliMerge Result}
		\label{subfig:ideal-fc1}
	\end{subfigure}
	
    \caption{\revision{A formatting conflict reported by \git and \ourtool along with the expected merge resolution by \intellimerge (deeplearning4j [\texttt{\href{https://github.com/Symbolk/deeplearning4j/commit/6c2853248afe9f0722fb2acad5551f687dc27e52}{6c285324}}]).} }
        \label{fig:formatting-conflict}
\vspace{-4mm}
\end{figure}

\revision{
\paragraph*{\textit{False Positives/Negatives \git Results}}
Table~\ref{tab:categories-git} shows the reasons behind the false positives and false negatives for Git.
There are generally three main reasons for false positives.
The most prevalent reason for \git's false positives is not being able to handle refactorings, and thus reporting conflicts that could be resolved automatically. 
There are \checkNum{140 (58\%)} false positive conflicts that \git reports that involve refactorings.
Given the selection of merge scenarios we use in our evaluation, it is natural to find that many of the conflicts \git reports are related to refactorings.
Table~\ref{tab:categories-git} also shows that \checkNum{61 (25\%)} of \git's reported false positives are due to ordering conflicts.
An \textit{ordering conflict} is a conflict caused by adding two program elements to the same location and the merge tool not knowing which order to put them in, when the order does not matter~\cite{Apel}.
Consider Figure~\ref{fig:ordering-conflict} where two import statements are added to the same location in \texttt{CompilationTestHelper.java}. 
\git does not know which order to put the new lines in, even though they can be put in any order. 
Finally, the remaining \checkNum{41 (17\%)} of \git's false positives are \textit{formatting conflicts} that are caused by different formatting between branches, such as additional white space or a new line on one branch that does not exist on the other. Figure~\ref{fig:formatting-conflict} shows a formatting conflict where the left branch does not have any space between the parameters while the right branch added a space.

}

\begin{figure*}[t]
	\centering
	\begin{subfigure}{.25\textwidth}
		\centering
		\includegraphics[width=\textwidth]{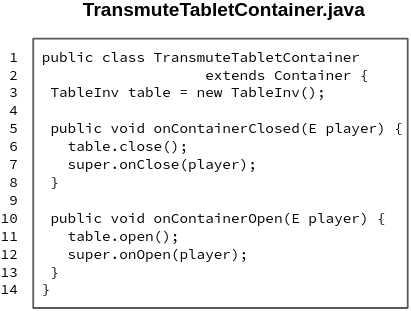}
		\caption{Base commit}
		\label{subfig:base-fn}
	\end{subfigure} \\
		\begin{subfigure}{.5\textwidth}
		\centering
		\includegraphics[width=\textwidth]{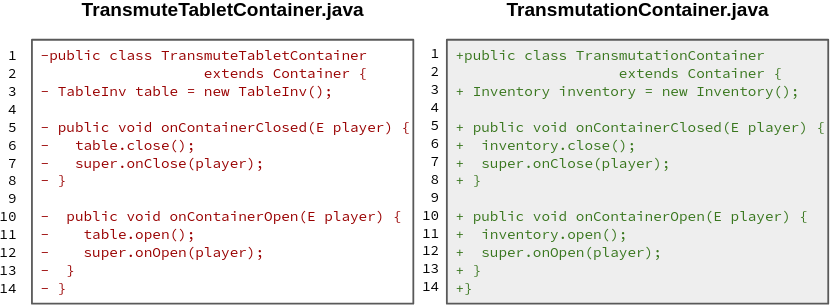}
		\caption{Left parent}
		\label{subfig:left-fn}
	\end{subfigure} 
		\begin{subfigure}{.25\textwidth}
		\centering
		\includegraphics[width=\textwidth]{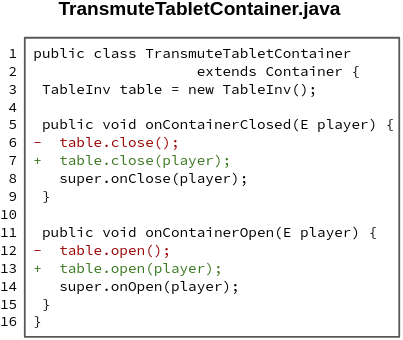}
		\caption{Right parent}
		\label{subfig:right-fn}
	\end{subfigure} 
	\begin{subfigure}{.5\textwidth}
		\centering
		\includegraphics[width=\textwidth]{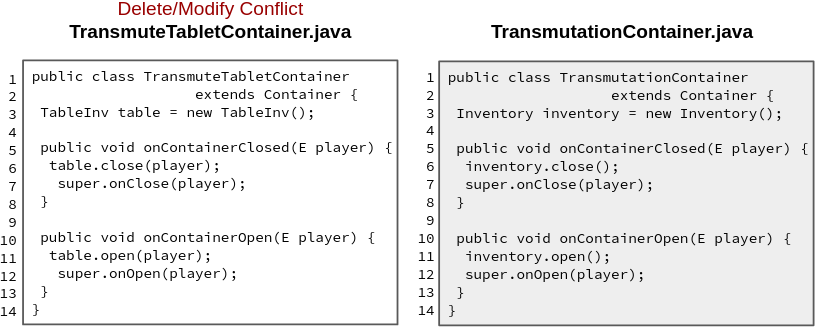}
		\caption{Git Result}
		\label{subfig:git-fn}
	\end{subfigure} 
		\begin{subfigure}{.25\textwidth}
		\centering
		\includegraphics[width=\textwidth]{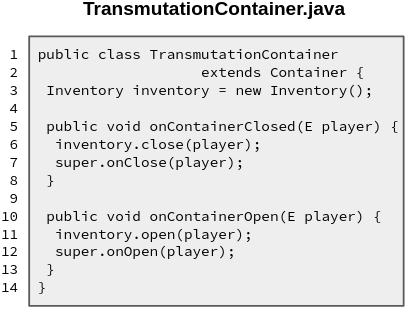}
		\caption{RefMerge Result}
		\label{subfig:ideal-fn}
	\end{subfigure}
	
\caption{\revision{The three versions (base, left, and right) of code involved in two of \git's false negatives, as well as the results merged by \git and \ourtool's result (ProjectE [\texttt{\href{https://github.com/sinkillerj/ProjectE/commit/12b0545e521f798a818fc37fc2560f8c92bca9ab}{12b0545e}}])
}.
}
\label{fig:git-merge-fn}
\vspace{-4mm}
\end{figure*}

\revision{
Not being able to handle refactorings also causes \git's results to include false negatives caused by syntax errors.
Figure~\ref{fig:git-merge-fn} presents a merge scenario where \git results in a false positive as well as false negatives.
On the left branch, file \texttt{TransmuteTabletContainer.java} is renamed to \texttt{TransmutationContainer.java} and class \texttt{TransmuteTabletContainer} is renamed to class \texttt{TransmutationContainer}. Additionally, class \texttt{TableInv} (which is the type of field \texttt{table} on Line 3) is renamed to \texttt{Inventory} as part of a \textit{Rename Class} refactoring. Note that \texttt{TableInv} is a class in \texttt{TableInv.java}, which we do not show in Figure~\ref{fig:git-merge-fn} for better visualization.
To match the type update,  field \texttt{table} is also renamed to \texttt{inventory}.
On the right branch, an \textit{Add Parameter} refactoring was performed on the definitions of methods \texttt{close} and \texttt{open} in \texttt{TableInv} (not shown), which resulted in the update of the function calls on Lines 6 and 12 of Figure~\ref{subfig:right-fn}.

We can see \git's merge result in Figure~\ref{subfig:git-fn} where it reports a delete/modify conflict in \texttt{TransmuteTabletContainer.java}, because it mistakenly sees that the left branch deleted the whole file (i.e., does not see this as a renamed class) while the right branch changed it.
This is false positive conflict, because the developer will need to deal with this reported conflict, when a tool that considers refactoring semantics can avoid this altogether. 
Note that when \git reports a delete/modify conflict, it does not physically delete the file because it waits for the developer's resolution.
\git also does not try to detect conflicts on the internal content of the file. 
Moreover, in \git's resolution, it shows the newly added file \texttt{TransmutationContainer.java} as is and does not provide any context that it is related to \texttt{TransmuteTabletContainer.java}.
Worse, \git's resolution of \texttt{TransmutationContainer.java} contains compilation errors, because the \texttt{close()} and \texttt{open()} calls are missing the added parameter.
This is a false negative.

Figure~\ref{subfig:ideal-fn} shows \ourtool's resolution, which illustrates the strengths of a refactoring-aware merging tool.
We can see that no modify/delete conflict is reported and that \texttt{TransmuteTabletContainer.java} is correctly deleted. 
In addition, field \texttt{table} is renamed to \texttt{inventory}, along with the rename of its type from \texttt{TableInv} to \texttt{Inventory}.
Although \ourtool does not support \textit{Add Parameter}, the added parameters are the only changes after \ourtool inverts \textit{Rename Class} and \textit{Rename Field}, resulting in the correct method calls on lines 6 and 11.
We note that in this scenario, \intellimerge incorrectly deletes classes \texttt{TableInv} and \texttt{TransmuteTabletContainer} instead of renaming them. 
In \intellimerge's result, files \texttt{TransmuteTabletContainer.java}, \texttt{TransmutationContainer.java}, \texttt{TableInv.java}, and \texttt{Inventory.java} do not exist, resulting in an incorrect merge.
}

\begin{table}[t]
\centering
\caption{The reason and frequency for false positives and false negatives reported by \ourtool.
}
\label{tab:categories-refMerge}
\hspace{0.1em}
\resizebox{0.45\textwidth}{!}{
\begin{tabular}{@{}llr@{}}
\toprule
\textbf{\ourtool Reasons} & \textbf{Type}     & \textbf{Frequency}       \\
\midrule 
\revision{Ordering Conflict}                   &   False Positive  &   61      \\
\revision{Formatting Conflict}                 &   False Positive  &   41      \\
\revision{Unsupported Refactoring}    &   False Positive  & 41        \\
\revision{Refactoring-related Formatting Conflict} &False Positive &   16      \\
\revision{Fails to Invert Refactoring}         &   False Positive  &   14       \\
\revision{IntelliJ Optimization}               &   False Positive  &   5       \\
\revision{Undetected Refactoring}              &   False Positive  &   5       \\
\revision{Refactoring-related Ordering Conflict} &  False Positive  &   5      \\
\midrule
\revision{Total}&                                      &188 \\
\bottomrule
\end{tabular}}

\vspace{-4mm}
\end{table}

\begin{figure}[t]
	\centering
		\begin{subfigure}{.225\textwidth}
		\centering
		\includegraphics[width=\textwidth]{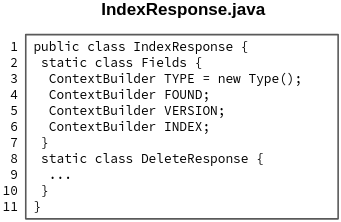}
		\caption{Base Commit}
		\label{subfig:refmerge-oc-base}
	\end{subfigure} \\
		\begin{subfigure}{.225\textwidth}
		\centering
		\includegraphics[width=\textwidth]{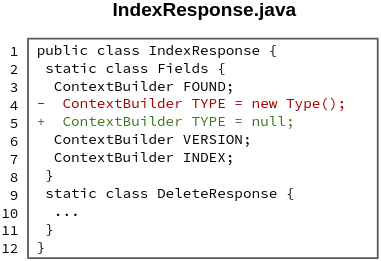}
		\caption{Left Commit}
		\label{subfig:refmerge-oc-left}
	\end{subfigure}
		\begin{subfigure}{.225\textwidth}
		\centering
		\includegraphics[width=\textwidth]{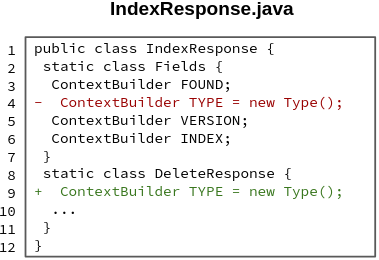}
		\caption{Right Commit}
		\label{subfig:refmerge-oc-right}
	\end{subfigure} \\
	\begin{subfigure}{.225\textwidth}
		\centering
		\includegraphics[width=\textwidth]{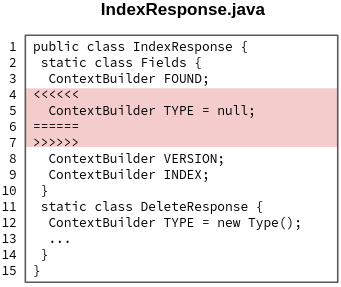}
		\caption{Git Result}
		\label{subfig:git-oc}
	\end{subfigure}
		\begin{subfigure}{.225\textwidth}
		\centering
		\includegraphics[width=\textwidth]{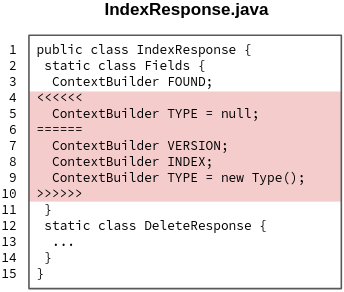}
		\caption{RefMerge Result}
		\label{subfig:refmerge-oc}
	\end{subfigure}
		\begin{subfigure}{.225\textwidth}
		\centering
		\includegraphics[width=\textwidth]{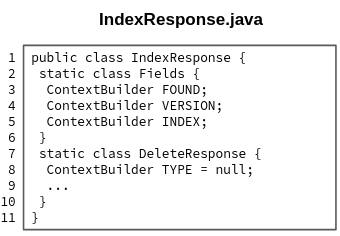}
		\caption{IntelliMerge Result}
		\label{subfig:intellimerge-oc}
	\end{subfigure}
	
\caption{\revision{The three versions (base, left, and right) of code involved in a refactoring-related ordering conflict that \ourtool introduces while along with \git's and \intellimerge's results (elasticsearch [\texttt{\href{https://github.com/elastic/elasticsearch/commit/503a166b7148a79ac6221894eca4835ef68d3480}{503a166b}}]).}}
\label{fig:refmerge-oc}
\vspace{-4mm}
\end{figure}

\revision{
\paragraph*{\textit{False positive/negative \ourtool Results}} Table~\ref{tab:categories-refMerge} shows the reasons behind the false positives and negatives for \ourtool. 
Similar to Git, \ourtool also suffers from being unable to resolve ordering and formatting conflicts, reporting the same \checkNum{61} ordering conflicts and \checkNum{41} formatting false positives as Git. 
\ourtool reports an additional \checkNum{5} ordering conflicts and \checkNum{16} formatting conflicts that arise from its refactoring handling, totalling \checkNum{66 (35\%)} ordering conflicts and \checkNum{55 (29\%)} formatting conflicts. 
All of the additional ordering conflicts are caused by move-related refactorings such as \texttt{Move Inner Class} or \texttt{Pull Up Method} being moved to the correct class but not being moved to the correct location within the file.
For example, Figure~\ref{fig:refmerge-oc} shows a merge scenario where \ourtool introduces a refactoring-related ordering conflict.
In this example, the left branch changes the field declaration \texttt{ContextBuilder Type = new Type()} to \texttt{ContextBuilder TYPE = null}.
The right branch moves field \texttt{TYPE} from inner class \texttt{IndexResponse.Fields} to inner class \texttt{IndexResponse.DeleteResponse}.
This results in \git reporting the conflict in Figure~\ref{subfig:git-oc} because \git sees that the left branch made a change to the line that the right branch deleted.
When \ourtool inverts the \textit{Move Field} refactoring, it correctly moves \texttt{TYPE} to \texttt{IndexResponse.Fields} but it moves it to the wrong location textually, moving it to line 9 instead of line 5.
If \ourtool inverts \texttt{TYPE} to the correct location, \ourtool would be able to resolve the conflict.
However, as shown in Figure~\ref{subfig:refmerge-oc}, \texttt{VERSION} and \texttt{INDEX} are unnecessarily part of the conflict while they originally were not.
While \ourtool complicates this conflict, \intellimerge resolves the conflict by moving \texttt{TYPE = null} to \texttt{IndexResponse.DeleteResponse}, as shown in Figure~\ref{subfig:intellimerge-oc}.

}

\begin{figure}[t]
	\centering
	\begin{subfigure}{.225\textwidth}
		\centering
		\includegraphics[width=\textwidth]{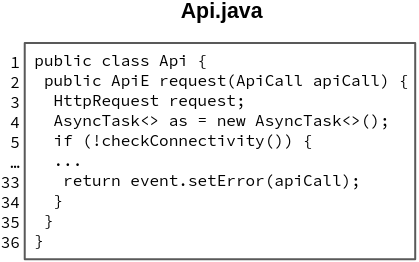}
		\caption{Base Commit}
		\label{subfig:refmerge-fc-base}
	\end{subfigure} \\
		\begin{subfigure}{.225\textwidth}
		\centering
		\includegraphics[width=\textwidth]{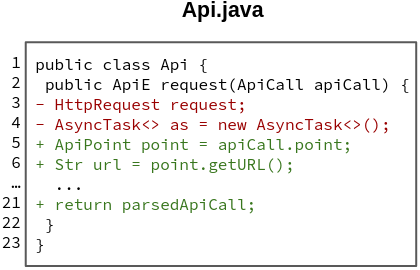}
		\caption{Left Commit}
		\label{subfig:refmerge-fc-left}
	\end{subfigure}
	\begin{subfigure}{.225\textwidth}
		\centering
		\includegraphics[width=\textwidth]{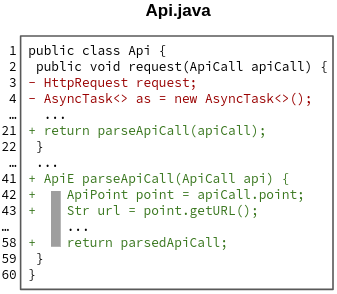}
		\caption{Right Commit}
		\label{subfig:refmerge-fc-right}
	\end{subfigure} \\
	\begin{subfigure}{.225\textwidth}
		\centering
		\includegraphics[width=\textwidth]{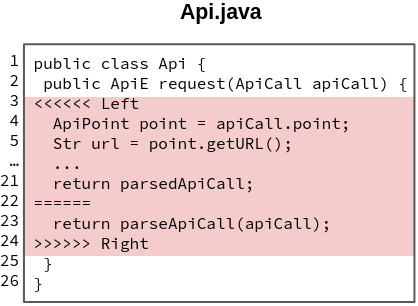}
		\caption{Git and IntelliMerge Result}
		\label{subfig:refmerge-fc-git}
	\end{subfigure}
	\begin{subfigure}{.225\textwidth}
		\centering
		\includegraphics[width=\textwidth]{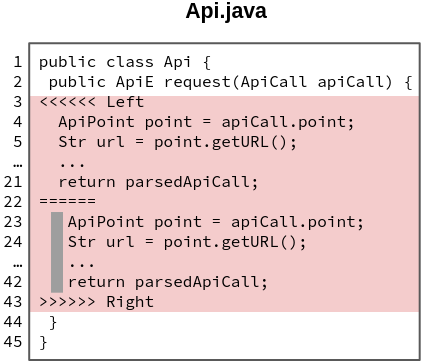}
		\caption{RefMerge Result}
		\label{subfig:refmerge-fc-rm}
	\end{subfigure} 
	
\caption{\revision{The three versions (base, left, and right) of code involved in a refactoring-related formatting conflict that \ourtool introduces along with \git's and \intellimerge's results (iFixitAndroid [\texttt{\href{https://github.com/iFixit/iFixitAndroid/commit/91932083f2f73d334f5789182da8117e275dc5f8}{91932083}}]).}}
\label{fig:refmerge-fc}
\vspace{-4mm}
\end{figure}

\revision{
The additional formatting conflicts are caused by formatting differences from inverting refactorings, also typically \texttt{Move Method} and \texttt{Move Inner Class} refactorings. 
In these conflicts, \ourtool resolves the refactoring conflict but leaves a formatting conflict that usually consists of two lines with different amounts of white space. 
Figure~\ref{fig:refmerge-fc}  provides an example where \ourtool results in a larger formatting conflict. 
In this example, the right branch makes the same changes to method \texttt{request} as the left branch, except the right branch also extracts the changes to method \texttt{parseApiCall}.
As shown in Figure~\ref{subfig:refmerge-fc-git}, \git reports a conflict involving \textit{Extract Method}, but this conflict
 has potential to be resolved by \ourtool and \intellimerge. 
Figure~\ref{subfig:refmerge-fc-rm} shows that \ourtool inverts the \textit{Extract Method} refactoring but has more space on each line than the right branch.
We mark the additional space on the right branch with gray to make the spacing difference clear.
The difference in spacing results in a formatting conflict twice as large as the conflict reported by \git.
It is worth noting that whenever \ourtool inverts an \textit{Extract Method} refactoring and it cannot resolve the conflict, \ourtool usually results in a conflict twice as large. 
While we typically associate larger conflicting regions with a more difficult merge resolution, we believe that \ourtool's process has its own merits. 
For example, while \git and \intellimerge report a smaller conflicting region than \ourtool, a developer trying to resolve the merge conflict first has to find the extracted method, then compare the changes between the two methods, and finally find the commit the method was extracted from to understand what the method looked like before the extract method refactoring. 
While \ourtool doubles the size of the conflict since it has the complete bodies of the two versions of the method (with and without the extracted code), the changes are shown side by side so a developer does not need to search for the extracted method.
Additionally, it is more clear what the method body changes were without the extracted method's parameters making things less clear. 
Thus, \ourtool actually provides additional information about the refactorings originally involved within the conflicting region to provide the context needed to resolve the conflict.

In \checkNum{41 (22\%)} of the false positives that \ourtool reports, the underlying issue is a refactoring that is not supported in the current implementation. For example, a merge conflict in MinecraftForge ([\texttt{\href{https://github.com/MinecraftForge/MinecraftForge/commit/f5781488d9da0d045cef802cf98af70e80eaa8aa}{f5781488}}]) contains an \textit{Add Parameter} refactoring and while \ourtool does resolve the conflict, \intellimerge does.

\ourtool reports \checkNum{14 (7\%)} false positives involving refactoring conflicts it supports but that it fails to resolve, because it could not invert the refactoring. \checkNum{Nine} of the refactorings that \ourtool fails to invert are \textit{Rename Method} refactorings.
After investigation, we found that these are typically caused by \ourtool being unable to find the refactored program element in IntelliJ's AST due to technical issues in our code, which we plan to fix.

There are \checkNum{five (3\%)} false positives caused by \textit{IntelliJ optimizations}, which are automatic optimizations done to the code after using the refactoring engine. All five of the IntelliJ optimizations were caused by inverting refactorings that were not involved in the original refactoring conflicts reported by \git. An example of this is replacing several import statements with \texttt{import package.*}, which then cause \git to detect a conflict in the merging step. 

Finally, there are \checkNum{five (3\%)} false positives that are due to \textit{undetected refactorings} that RefactoringMiner did not detect. In these scenarios, there are several methods that are similar, both in structure and naming, which likely made it difficult for RefactoringMiner to detect the refactoring. We reported the issue to the RefactoringMiner developers.

}

\begin{table}[t]
\centering
\caption{The reason and frequency of  false positives and false negatives reported by \intellimerge.
}
\label{tab:categories-intelliMerge}
\hspace{0.1em}
\resizebox{0.4\textwidth}{!}{
\begin{tabular}{@{}llr@{}}
\toprule
\textbf{\intellimerge Reasons} & \textbf{Type}     & \textbf{Frequency}       \\
\midrule 
\revision{Matching Error}           &   False Positive      &   646         \\
\revision{Undetected Refactoring}      &   False Positive  &   37  \\
\revision{Ordering Conflict}           &   False Positive      &   10          \\
\revision{Incorrectly Detected Refactoring}         &   False Positive  &   6  \\
\revision{Unsupported Refactoring}  &   False Positive &    5 \\
\revision{Formatting Conflict}         &   False Positive      &   3   \\
\revision{Deletes Conflict Block}     &   False Negative      &   45  \\
\revision{Matching Error}              &   False Negative      &   21 \\
\revision{Incorrectly Detected Refactoring}        &   False Negative  &   5  \\

\midrule
\revision{Total}&                                      & 778 \\
\bottomrule
\end{tabular}}
\end{table}


\revision{
\paragraph*{\textit{False positive/negative \intellimerge Results}}
Table~\ref{tab:categories-intelliMerge} shows the reasons behind the false positives and negatives for \intellimerge.
We start with some of the reasons we already observed for the other tools.
\intellimerge reports \checkNum{10 (1\%)} false positives due to ordering conflicts and also has \checkNum{37 (5\%)} because of undetected refactorings. \intellimerge also sometimes fails to detect a refactoring, most commonly with parameter-level refactorings \checkNum{(14)}, \textit{Extract Method} (9), and \texttt{Rename Class} \checkNum{(5)}.
The undetected refactorings can be split into two groups: (1) the presence of several similar program elements drops the correct refactored program element below the similarity threshold, and (2) the presence of several changes to a program element cause \intellimerge to think that the refactored program element is an addition.
For example, \intellimerge missed the \textit{Extract Method} refactoring in Figure~\ref{fig:refmerge-fc} that \ourtool was able to detect. In this specific scenario, several methods had \texttt{parse} or \texttt{ApiCall} in their name, resulting in an incorrect match.
In addition, there are 125 refactorings that were performed in class \texttt{Api}, resulting in several changes that made it more difficult for \intellimerge to detect the refactoring.
}

\begin{figure}[t]
	\centering
	\begin{subfigure}{.225\textwidth}
		\centering
		\includegraphics[width=\textwidth]{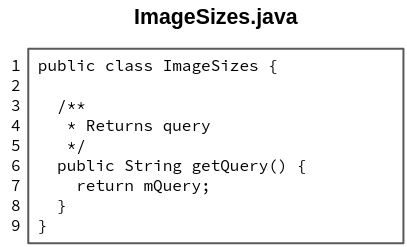}
		\caption{Git and RefMerge Result}
		\label{subfig:git-matching}
	\end{subfigure}
		\begin{subfigure}{.225\textwidth}
		\centering
		\includegraphics[width=\textwidth]{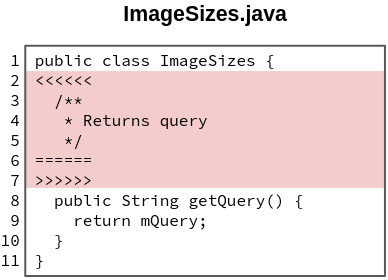}
		\caption{IntelliMerge Result}
		\label{subfig:intellimerge-matching}
	\end{subfigure}
	
\caption{\revision{A conflict reported by \intellimerge caused by a matching error along with \git's result (iFixitAndroid [\texttt{\href{https://github.com/iFixit/iFixitAndroid/commit/91932083f2f73d334f5789182da8117e275dc5f8}{91932083}}]).}}
\label{fig:matching-conflict}
\vspace{-4mm}
\end{figure}

\revision{
Note that, unlike \git and \ourtool, \intellimerge reports only \checkNum{three} false positives related to formatting conflicts.
On the other hand, \checkNum{646} of \intellimerge's false positives (\checkNum{91\%}) are due to matching errors.
We define a \textit{matching error} as an error caused by \intellimerge's graph node matching process. This primarily happens with comments, annotations, and imports where \intellimerge cannot find matches for these nodes and assumes they were deleted. 
In Figure~\ref{fig:matching-conflict}, \git does not report a merge conflict. As seen in Figure~\ref{subfig:intellimerge-matching}, \intellimerge incorrectly reports a conflict that indicates that the comment was changed on one branch and deleted on the other. 
In this example, the left branch added method \texttt{getQuery} along with \texttt{getQuery}'s comment to class \texttt{ImageSizes}.
The right branch does not add any changes to class \texttt{ImageSizes}.
While conflicts caused by matching errors are typically small, they happen frequently and the developer needs to spend time to investigate the conflict and decide that they can ignore it.

}

\begin{figure}[t]
	\centering
	\begin{subfigure}{.225\textwidth}
		\centering
		\includegraphics[width=\textwidth]{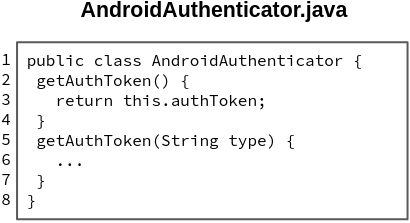}
		\caption{Base Commit}
		\label{subfig:im-mc-fp-base}
	\end{subfigure} \\
		\begin{subfigure}{.225\textwidth}
		\centering
		\includegraphics[width=\textwidth]{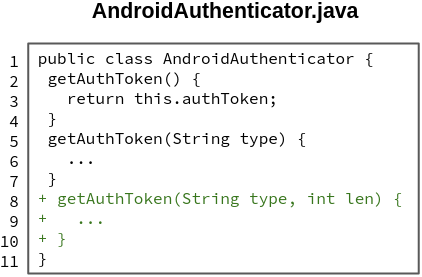}
		\caption{Left Commit}
		\label{subfig:im-mc-fp-left}
	\end{subfigure}
	\begin{subfigure}{.235\textwidth}
		\centering
		\includegraphics[width=\textwidth]{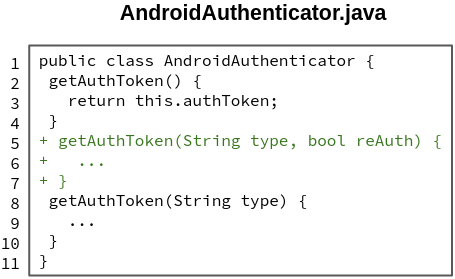}
		\caption{Right Commit}
		\label{subfig:im-mc-fp-right}
	\end{subfigure} \\
	\begin{subfigure}{.225\textwidth}
		\centering
		\includegraphics[width=\textwidth]{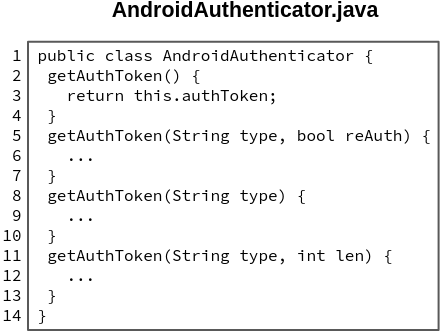}
		\caption{Git and RefMerge Result}
		\label{subfig:im-mc-fp-git}
	\end{subfigure}
	\begin{subfigure}{.225\textwidth}
		\centering
		\includegraphics[width=\textwidth]{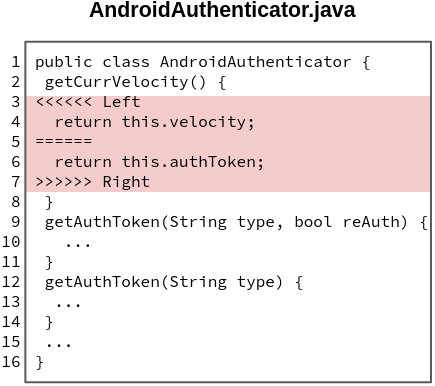}
		\caption{IntelliMerge Result}
		\label{subfig:im-mc-fp-im}
	\end{subfigure}

\caption{\revision{The three versions (base, left, and right) of code involved in a conflict reported by \intellimerge caused by incorrectly detecting a refactoring along with the results from \git and \ourtool (platform\_fmwk\_support [\texttt{\href{https://github.com/aosp-mirror/platform_frameworks_support/commit/6c371fc5da43c7841549d0b7573831ca7dd1ca9f}{6c371fc5}}]).} }
\label{fig:ref-conf-fp}
\vspace{-4mm}
\end{figure}

\revision{
\checkNum{Six} false positives are caused by \intellimerge incorrectly detecting a refactoring that was never performed. There are \checkNum{six} cases where \intellimerge incorrectly matches methods in different classes, resulting in \intellimerge incorrectly moving a method and reporting a conflict in its method body. In Figure~\ref{fig:ref-conf-fp}, \git does not report a merge conflict.
As shown in Figure~\ref{subfig:im-mc-fp-left}, the developers add method \texttt{getAuthToken(String type, int len)} to \texttt{AndroidAuthenticator.java}.
On the right branch, the developers add method \texttt{getAuthToken(String type, bool reAuth)} to \texttt{AndroidAuthenticator.java}.
The right branch also adds method \texttt{getCurrVelocity} to class \texttt{ScrollerCompat.java}, not shown in this figure.
As shown in Figure~\ref{subfig:im-mc-fp-git}, both \git and \ourtool do not report a conflict. 
Figure~\ref{subfig:im-mc-fp-im} shows the result for \intellimerge.
\intellimerge deletes \texttt{getCurrVelocity} in \texttt{ScrollerCompat.java} and replaces \texttt{getAuthToken()} in \texttt{AndroidAuthenticator} with \texttt{getCurrVelocity()}. 
\intellimerge reports a conflict in \texttt{getCurrVelocity} with the changes added to \texttt{getCurrVelocity} on the left side and the already existing method body of method \texttt{getAuthToken} on the right, resulting in a false positive.

The remaining \checkNum{5} false positives are caused by unsupported refactorings. Both \intellimerge and \ourtool do not support \textit{Extract Superclass}. 
}

\begin{figure}[t]
	\centering
	\begin{subfigure}{.225\textwidth}
		\centering
		\includegraphics[width=\textwidth]{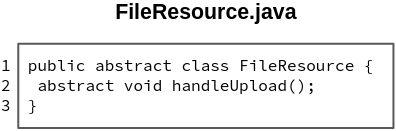}
		\caption{Git and RefMerge Result}
		\label{subfig:git-syntax-fn}
	\end{subfigure}
		\begin{subfigure}{.225\textwidth}
		\centering
		\includegraphics[width=\textwidth]{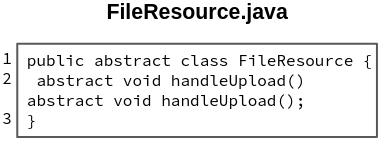}
		\caption{IntelliMerge Result}
		\label{subfig:im-syntax-fn}
	\end{subfigure}
	
\caption{\revision{A syntax error (Duplicate declaration) that \intellimerge's merge resolution introduces versus \git's and \ourtool's result (deeplearning4j [\texttt{\href{https://github.com/Symbolk/deeplearning4j/commit/8d1ff15ffcdbf5901db43226a082d5d86f617e15}{8d1ff15f}}]).}}
\label{fig:intellimerge-syntax-fn}
\vspace{-4mm}
\end{figure}

\begin{figure}[t]
	\centering
	\begin{subfigure}{.225\textwidth}
		\centering
		\includegraphics[width=\textwidth]{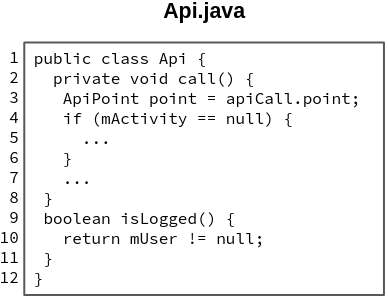}
		\caption{Base commit}
		\label{subfig:base-im-incorrect}
	\end{subfigure} \\
		\begin{subfigure}{.225\textwidth}
		\centering
		\includegraphics[width=\textwidth]{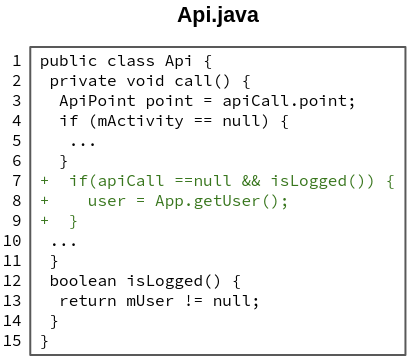}
		\caption{Left parent}
		\label{subfig:left-im-incorrect}
	\end{subfigure} 
		\begin{subfigure}{.225\textwidth}
		\centering
		\includegraphics[width=\textwidth]{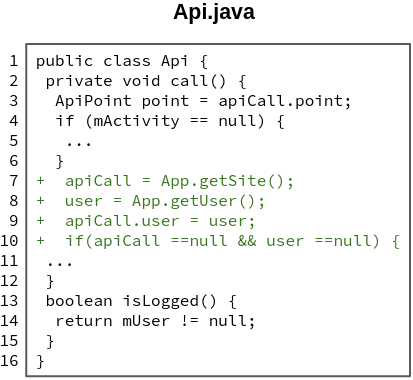} 
		\caption{Right parent}
		\label{subfig:right-im-incorrect}
	\end{subfigure} \\
	\begin{subfigure}{.225\textwidth}
		\centering
		\includegraphics[width=\textwidth]{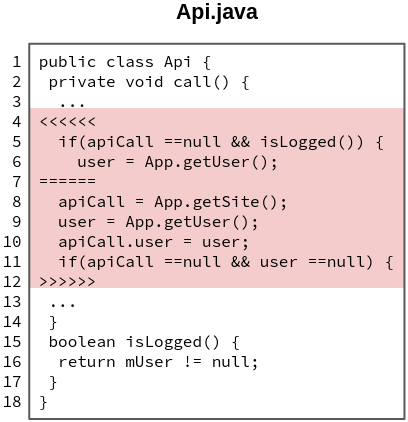}
		\caption{Git and RefMerge Result}
		\label{subfig:git-im-incorrect}
	\end{subfigure} 
		\begin{subfigure}{.225\textwidth}
		\centering
		\includegraphics[width=\textwidth]{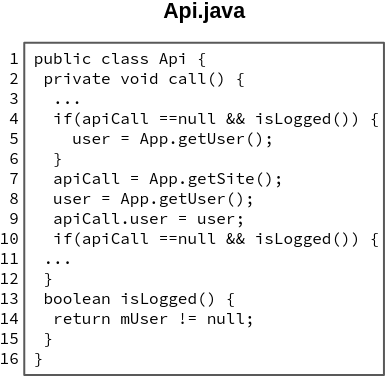}
		\caption{IntelliMerge Result}
		\label{subfig:im-incorrect}
	\end{subfigure}
	
\caption{\revision{The three versions (base, left, and right) of code involved in a necessary conflict that \git and \ourtool report and \intellimerge misses due to incorrectly detecting a refactoring (iFixitAndroid [\texttt{\href{https://github.com/iFixit/iFixitAndroid/commit/91932083f2f73d334f5789182da8117e275dc5f8}{91932083}}]).}}
\label{fig:im-incorrect-ref}
\vspace{-4mm}
\end{figure}

\revision{
\intellimerge results in \checkNum{45 (63\%)} false negatives because it incorrectly deletes the changes made in the left and right branches, causing
it to completely delete code that should have resulted
in conflict blocks.
We find that \intellimerge frequently incorrectly deletes complete classes that are involved in \textit{Rename Class} or \textit{Move Class} refactorings, such as when it deleted \texttt{TransmuteTabletContainer.java} in the example in Figure~\ref{fig:git-merge-fn}.
This is likely caused by a matching error where if \intellimerge cannot find a match for program elements in the base commit, it assumes these elements were deleted and removes them accordingly.
Note that this is a lower bound for how many times \intellimerge could have deleted other files or program elements that were not part of a conflict block and since we focus on the reported conflicts by each tool, we would have missed this happening in files where no merge tool reports a conflict.
For example, there is a merge scenario in MinecraftForge ([\texttt{\href{https://github.com/MinecraftForge/MinecraftForge/commit/c3559b2dbdc6462f496d606e25bf081920b181f9}{c3559b2d}]}) where \intellimerge deletes every file in the scenario except for one, including six files that should contain conflicts.

We find that \checkNum{21 (30\%)} of \intellimerge's false negatives are due to matching errors that eventually lead to syntax errors. Most of the syntax errors seem to happen in classes that contain several method-level refactorings and several similar method declarations. 
Figure~\ref{fig:intellimerge-syntax-fn} shows a merge scenario where \intellimerge results in a duplicate declaration syntax error while \git and \ourtool do not. In this example, neither branch changed method \texttt{handleUpload} on line 2 of \texttt{FileResource.java}. However, \intellimerge's resolution results in a syntax error where it duplicates the method declaration on line 2.

Finally, the \checkNum{5 (7\%)} remaining false negatives are due to \intellimerge detecting refactorings that were not performed, leading to \intellimerge moving methods to classes that the developers never moved them to and causing additional syntax errors.
This is similar to the false positives caused by incorrectly detected refactorings.
However, in this case \intellimerge does not report the expected conflict block. 
Figure~\ref{fig:im-incorrect-ref} provides an example where \intellimerge misses a conflict. 
In this example, the left branch and right branch add new code to the same spot in method \texttt{call}.
As shown in Figure~\ref{subfig:git-im-incorrect}, \git and \ourtool report a conflict with the conflicting region containing the additions from each branch.
This is a necessary conflict because the left branch sets \texttt{user} in the if statement, while the right branch sets \texttt{user} before the if statement.
Additionally, method \texttt{isLogged} checks if \texttt{mUser != null} instead of \texttt{user == null}.
In Figure~\ref{subfig:im-incorrect}, \intellimerge incorrectly replaces \texttt{user == null} with \texttt{isLogged()} on line 10 even though method \texttt{isLogged} returns \texttt{mUser != null}.
Thus resulting in a different logic check than is expected and resulting in a false negative.
}

\subsection{Interpretation of RQ2 Results}

\revision{
The above results indicate that, in our sample, \ourtool automatically resolves some of \git's reported conflicts, which results in less false positives than \git \checkNum{~(188 versus 243)}. On the other hand, \intellimerge almost triples the number of false positives \checkNum{(707 versus 242)}. 
While \intellimerge reports more false positives than \git and \ourtool, \intellimerge does well with ordering and formatting conflicts due to its graph-based approach. \intellimerge also decreases the number of refactoring conflicts a developer needs to deal with, but this comes at the price of many more false positives: \checkNum{646} of \intellimerge's false positives are matching errors which are typically small in size. This explains the quantitative results of RQ1 where \intellimerge reports more conflicts but less conflicting LOC.
Additionally, while \intellimerge does not detect refactorings in \checkNum{37} conflict blocks reported by \git (struggling most with parameter level refactorings), \intellimerge typically does well with the refactoring conflicts it does detect.
However, \intellimerge also incorrectly detects \checkNum{11} refactorings (six false positives and five false negatives from Table~\ref{tab:categories-intelliMerge}) and reports a total of \checkNum{71} false negatives.
Thus, while the results of RQ1 show that \intellimerge works well for a small proportion of scenarios where it is able to highly reduce the resulting conflicts in a scenario, our qualitative results suggest that some of these may actually be false negatives.

On the other hand, \ourtool does not miss any conflicts that need to be reported (i.e., completely eliminates false negatives) and reduces the number of false positives reported by \checkNum{23\%}, when compared to \git.
\ourtool worsens the situation at a much lower rate than \intellimerge, reporting \checkNum{26} false positives that were not reported by \git, where \checkNum{16} of these are formatting conflicts that are typically introduced after resolving a refactoring conflict.
In general, \ourtool struggles most with move-related refactorings where it introduces ordering conflicts.
}

\begin{findingenv}{RQ2 Summary}{finding:rq2}
\revision{
Compared to \git, \ourtool reduces the number of false positives by \checkNum{23\%} and completely eliminates false negatives while \intellimerge increases them by \checkNum{192\%} and \checkNum{1,320\%} respectively.
\ourtool struggles most with move-related refactorings whereas \intellimerge struggles most with parameter-level and class-level refactorings. 
}
\end{findingenv}

\section{Discussion}

In our study, we compared two refactoring-aware merging approaches that have not been compared before. 
\revision{RQ1 results show that \ourtool manages to resolve about twice as many conflicting merge scenarios as \intellimerge.
We found that while \intellimerge reduced the number of conflicting LOC in more scenarios compared to \ourtool, \intellimerge also increased the number of conflicting LOC in more scenarios. On the other hand, \ourtool makes the situation worse in a smaller proportion of merge scenarios and by a lower percentage increase.
Additionally, our qualitative analysis shows that \intellimerge reported a much higher number of false positives and false negatives whereas \ourtool reduced the number of reported false positives and completely eliminated false negatives in the sampled 50 merge scenarios.
Thus, operation-based refactoring-aware merging shows promise to help improve the developers' experience without the risk of increasing the number of false negatives. 
}


\paragraph*{\textit{Strengths and Weaknesses of \intellimerge}}
The nature of \intellimerge's graph-based approach makes it avoid formatting and ordering conflicts. 
However, \intellimerge seems to struggle with correctly matching graph nodes across the two versions of the code.
We believe that \intellimerge's use of a similarity score for its refactoring detection is one of the main reasons for this. 
\intellimerge often failed to detect a refactoring because the refactored program element was too similar to other existing program elements.
We also found cases where a non-refactoring change caused a program element to be within the similarity threshold of other program elements, causing \intellimerge to treat it as a refactoring. 
Although \intellimerge could potentially change the used similarity threshold, the use of a similarity score will always run into these problems~\cite{RefactoringMiner1}. 
\revision{
Additionally, we found that \intellimerge generally struggles with \textit{Rename Class} and \textit{Move Class} refactorings.
When class-level refactorings are performed, \intellimerge frequently deletes the entire related class.
However, \intellimerge seemed to do well with the other refactorings it detected.
}

\paragraph*{\textit{Strengths and Weaknesses of \ourtool}}

Whereas \intellimerge's graph-based approach makes it avoid formatting and ordering conflicts, \ourtool's operation-based approach is more prone to such conflicts. 
While formatting conflicts are a small price to pay considering they are typically easier to resolve than refactoring conflicts, move-related refactorings proved to be conceptually challenging when it comes to undoing/redoing them. 
Although \ourtool can move the program element to the correct class, it cannot guarantee that it is moved to the same textual location it was previously at.
\revision{
Despite this, \ourtool resolves or simplifies more refactoring-related conflicts than the complications it introduces, all while avoiding syntax errors.
Additionally, while the number of refactorings in a merge scenario can cause problems for \intellimerge, \ourtool is resilient to the number of refactorings in a given scenario.
}

\paragraph*{\textit{Moving Forward}}

Driven by these findings, we propose a few paths moving forward in refactoring-aware merging. 
We believe that improvements in graph-based refactoring-aware merging requires addressing the matching algorithm.
The current merging algorithm \intellimerge uses seems to work well, but the initial matching phase can be improved by avoiding the similarity score matching and instead using a refactoring detection algorithm such as that used in RefactoringMiner~\cite{RefactoringMiner1}.

\revision{We believe that \ourtool showed very promising results, despite supporting a subset of the refactorings \intellimerge supports. Future work could go in two different directions: 
(1) adding support for more refactoring types, and
(2) using language semantics to address ordering-related conflicts when possible, such as the approach proposed by Apel et al.~\cite{Apel}.

}


Finally, it could make sense to combine the two refactoring-aware approaches in some way similar to how changing strategies/auto-tuning between semi-structured and structured merge was previously proposed~\cite{JDime}. As the nature of graph-based merging seems to do well with ordering conflicts and formatting conflicts, this would address the weaknesses of operation-based merging. However, addressing the weaknesses caused by \intellimerge's matching algorithm would need to happen before this path could be considered further.

\section{Threats to Validity}

We explain the potential threats to the validity of our results.

\paragraph*{\textit{Construct Validity}} 


In our qualitative analysis, we manually compare the results of the three tools to identify false positives and false negatives.
This means we may miss false negatives that all three tools fail to report. 
Additionally, the analysis was done by a single author and is thus subject to their understanding of the scenario.
To alleviate this as much as possible, we compare the changes in the left parent, right parent, and base commit for each merge scenario to first try to understand the developer's intentions and the expected merge result.
We record a detailed description of our interpretation of the scenario and conflicts and share this in our artifact to allow further external validation.
Further analysis involving investigating run time and compile time errors could also further shed further light on false negatives reported by the three approaches.

\paragraph*{\textit{Internal Validity}} Any problems inherited from the tools used in \ourtool or in our evaluation setup may lead to inaccuracies in the results. To mitigate this, we carefully consider the role of each tool used in our study and analyze its results through manual verification.
While not a bug with IntelliJ per se, our qualitative analysis showed that IntelliJ's refactoring engine, which we use to invert and replay refactorings, performs optimizations that lead to unnecessary merge conflicts.
This means that the reported number of conflicts in our results is an upper bound and with engineering effort and help from the IntelliJ developers to allow us to disable these optimizations, these limitations can be mitigated. Alternatively, a different refactoring engine that does not force these optimizations can be used.
Any refactoring that RefactoringMiner misses will not be inverted and replayed, which will result in the same merge as Git. Any refactorings that RefactoringMiner detects which were not performed will result in RefMerge inverting and replaying a "fake" refactoring, which may lead to unnecessary merge conflicts. During our development, we came across some such occurrences and the RefactoringMiner author fixed these in the tool. In our qualitative analysis, we came across three refactorings that RefactoringMiner did not detect, which we recently reported.
Overall, RefactoringMiner achieves a precision of 98\% and 87\% recall~\cite{RefMiner}.
In general, it is important for \ourtool to rely on a tool with high precision to ensure we do not result in unnecessary conflicts.
A lower recall simply means \ourtool will result in the same resolution as \git.

\paragraph*{\textit{External Validity}} By selecting sample projects with different sizes and refactoring histories, we try our best to have a representative evaluation. Our evaluation is limited to Java open-source projects since both tools are Java specific. 
That said, while our implementation of \ourtool is Java specific, an operation-based approach does not need to be. 
Our qualitative analysis is based only on a sample of 50 merge scenarios due to the time consuming nature of the process (avg. 63min/scenario). However, the 50 merge scenarios we investigated have more than 1,000 unique merge conflicts. 
As far as we are aware, this is the most extensive qualitative analysis performed in terms of unique merge conflicts~\cite{Cavalcanti, JDime, Seibt}.
Naturally, investigating additional merge scenarios could reveal more for each tool.

\section{Related Work}

\paragraph*{\textit{\textbf{Software Merging}}}

The proposed software merging techniques in the literature can generally be categorized into \textit{unstructured}, \textit{structured}, and \textit{semi-structured} merging techniques~\cite{Mens}.


\textit{Unstructured merging techniques} represent any software artifact as a sequence of text lines~\cite{Berlin}. This gives unstructured merging techniques the strength of being able to process all textual artifacts, regardless of the programming language~\cite{Mens}. The downside to this technique is that unstructured merging cannot handle multiple changes to the same lines, since it cannot consider the syntactic and semantic meaning in software artifacts~\cite{Apel}.
Due to its simplicity and versatility, modern version-control systems such as \git or mercurial still rely on such unstructured merging.

\textit{Structured merging} tries to alleviate the problems of unstructured tools by leveraging the underlying structure of software artifacts, typically through operating on an Abstract Syntax Tree (AST) instead of textual lines~\cite{struct1}. Considering the structure of software artifacts allows structured merging techniques to handle syntactic and semantic conflicts~\cite{SemDiff, Zhu, Lebenich}. This comes at the cost of generally being language specific and being too expensive to be used in practice. 
\texttt{JDime} is a structured merge tool that is capable of tuning the merging process by switching between unstructured merge and structured merge~\cite{JDime}. Zhu et al.~\cite{AutoMerge} built on top of \texttt{JDime} by matching nodes based on an adjustable quality function. Leßenich et al.~\cite{lessenich2015balancing} proposed \textit{auto-tuning}, an approach that switches between structured and unstructured merging, and implemented \texttt{JDime} to demonstrate their approach. Seibt et al.~\cite{Seibt} recently performed a large-scale empirical study with unstructured, semi-structured, and structured merge algorithms and their findings suggest that combined strategies are promising moving forward.

\textit{Semi-structured techniques} aim to create a middle ground by considering both the language independence of unstructured merging and the precision of structured merging~\cite{Apel}. \texttt{FSTMerge} was proposed by Apel et al.~\cite{Apel} as one of the first semi-structured merging approaches. While \texttt{FSTMerge} reduces the number of merge conflicts reported compared to unstructured merge, \texttt{FSTMerge} struggles with renamings. Cavalcanti et al.~\cite{Cavalcanti} proposed \texttt{jFSTMerge}, building upon \texttt{FSTMerge} by adding handlers for different types of conflicts such as renaming.

By representing software artifacts partly as text and partly as trees, semi-structured merging achieves a certain level of language-independence. Cavalcanti et al.~\cite{Cavalcanti2} performed an empirical study to compare unstructured and semi-structured merging techniques. They found that semi-structured merge can reduce the number of merge conflicts by half. 
We compare only against \intellimerge because in their paper, they show that they outperform \texttt{jFSTMerge}~\cite{IntelliMerge}. Furthermore, we are focusing on techniques that specifically target refactorings in order to compare their strengths and weaknesses.

\paragraph*{\textbf{\textit{Proactive Conflict Detection \& Prevention}}}
The key idea behind this research line is that detecting conflicts as soon as they happen, even before a developer commits the changes, can lead to conflicts that are easier to resolve.
Knowing what changes other developers are making is beneficial for team productivity and reducing the number of reported merge conflicts~\cite{estler2014awareness}. One such approach is \textit{speculative merging}~\cite{Brun, Guimaraes}, where all combinations of available branches are pulled and merged in the background. 
Owhadi-Kareshk et al.~\cite{PMC} designed a classifier for predicting merge conflicts with the aim of reducing the computational costs of speculative merging by filtering out merge scenarios that are unlikely to be conflicting.

\texttt{Syde}~\cite{hattori2010syde} and \texttt{Palantir}~\cite{sarma2011palantir} are two tools that increase developer awareness by illustrating the code changes their team members are making. \texttt{Cassandra}~\cite{kasi2013cassandra} minimizes simultaneous edits to the same file by optimizing task scheduling.  \texttt{ConE}~\cite{ConE} is an approach that proactively detects concurrent edits to help mitigate certain resulting problems, including merge conflicts.
\revision{
Dewan et al. propose CollabVS, a semi-synchronous detection and resolution tool that detects a potential conflict when a user starts editing a program element that has a dependency on another program element that has been edited but not committed by another developer~\cite{Dewan2007}.
}
Silva et al.~\cite{Silva2} proposed utilizing automated unit test creation to detect semantic conflicts that a merge tool could have missed.
Fan et al.~\cite{Fan} proposed using dependency-based automatic locking to support fine-grained locking and avoid semantic conflicts. 
DeepMerge is a recent effort that defines merge conflict resolution as a machine learning problem~\cite{DEEPMERGE}. The approach primarily leverages the fact that around 80\% of merge conflict resolution only rearrange lines~\cite{Rearrange}. However, they do not explicitly consider refactoring semantics in their merge conflict resolution.

\paragraph*{\textit{\textbf{Refactoring Detection}}}

Refactoring is a widespread practice that enables developers to improve the maintainability and readability of their code~\cite{Silva}. Refactoring has been extensively studied over the past few decades~\cite{Mens2}, with recent work focusing on the detection of refactoring changes and the relationship between refactorings and code quality~\cite{Choi,Palomba,Bavota,RIPE}. 
Multiple tools have been developed to detect different refactoring types, such as \texttt{Ref-Finder}~\cite{Kim} and \texttt{RefDistiller}~\cite{Alves}. We use the state-of-the-art refactoring detection tool, RefactoringMiner, which achieves a 
precision of 98\% and a recall of 87\%~\cite{RefMiner}.

\paragraph*{\textbf{\textit{Operation-based \& Refactoring-aware Merging}}}

Operation-based merging is a semi-structured merging technique that models changes between versions as operations or transformations~\cite{Edwards, Lie, Lippe} which could be used to support refactoring-aware merging~\cite{MolhadoRef}. Nishimura et al.~\cite{Nishimura} proposed a tool that reduces the manual effort necessary to resolve merge conflicts by replaying fine-grained code changes related to conflicting class members. Their approach only considers edits and has problems with long edit histories and finer granularity of operations~\cite{MolhadoRef}. 

\revision{
Similar to MolhadoRef, Ekman and Asklund~\cite{Ekman} 
present a refactoring-aware versioning system. Their approach is more lightweight since it keeps the program elements and their IDs in volatile memory, thus allowing for a
short-lived history of refactored program entities.
However, their approach has only implemented rename- and move-refactorings. 
Thus, it remains to be seen whether the approach is generic enough for other refactorings.
Additionally, \ourtool does not rely on short-lived history because it detects refactorings at any point in the life time of branches from the common ancestor until the merging point.
}

Dig et al.~\cite{MolhadoRef} proposed MolhadRef, an operation-based refactoring-aware merging algorithm that treats refactorings as operations and considers their semantics, and Shen et al.~\cite{IntelliMerge} proposed \intellimerge, a graph-based refactoring-aware algorithm. Since we discuss these two approaches in detail in the paper and in our empirical comparison, we do not discuss them again here.

\section{Conclusion}

In modern software development, version control systems play a crucial role in enabling developers to collaborate on large projects.
Most modern version control systems use unstructured merging techniques that do not understand code-change semantics.
\revision{
{In this paper, we rejuvenate operation-based refactoring-aware merging~\cite{MolhadoRef} with the hope that this invigorates a fruitful line of research that has a large potential for practical impact.}
We design and implement the first \git-based refactoring-aware merging implementation in \ourtool. We add support for 17 refactoring types, including \textit{Extract Method} and \textit{Inline Method} which were argued to be a limitation for operation-based merging~\cite{IntelliMerge}. 
We perform the first large-scale empirical evaluation of operation-based refactoring-aware merging, implemented in \ourtool, and compare it to \intellimerge, a graph-based refactoring-aware merging technique~\cite{IntelliMerge}. Our evaluation on 2,001 merge scenarios from 20 open-source projects sheds light on the strengths and weaknesses of each approach.}
\revision{
We find that \ourtool is able to completely resolve or reduce the number of conflicts in more scenarios than \intellimerge without creating as much extra work for the developers.
We plan to explore two directions to further improve operation-based refactoring-aware merging:
(1) resolve refactoring-related ordering conflicts that our implementation of operation-based merging causes and 
(2) add support for more refactoring types.
}

\bibliographystyle{IEEEtran}
\bibliography{bibFile}

\end{document}